\documentclass[twocolumn]{aastex631}

\graphicspath{{./}{figures/}}
\usepackage{ amssymb }
\usepackage{ here }
\begin{document}

\title{Size - Stellar Mass Relation and Morphology of Quiescent Galaxies at $z\geq3$ in Public $JWST$ Fields}

\correspondingauthor{Kei~Ito}
\author[0000-0002-9453-0381]{Kei~Ito}
\altaffiliation{JSPS Research Fellow (PD)}
\email{kei.ito@astron.s.u-tokyo.ac.jp, kei.ito.astro@gmail.com}
\affiliation{Department of Astronomy, School of Science, The University of Tokyo, 7-3-1, Hongo, Bunkyo-ku, Tokyo, 113-0033, Japan}

\author[0000-0001-6477-4011]{Francesco Valentino}
\affiliation{European Southern Observatory, Karl-Schwarzschild-Str. 2, D-85748 Garching bei Munchen, Germany}
\affiliation{Cosmic Dawn Center (DAWN), Denmark}

\author{Gabriel Brammer}
\affiliation{Cosmic Dawn Center (DAWN), Denmark}
\affiliation{Niels Bohr Institute, University of Copenhagen, Jagtvej 128, DK-2200 Copenhagen N, Denmark}

\author[0000-0002-9382-9832]{Andreas L. Faisst}
\affiliation{Caltech/IPAC, 1200 E. California Blvd. Pasadena, CA 91125, USA}

\author[0000-0001-9885-4589]{Steven Gillman}
\affiliation{Cosmic Dawn Center (DAWN), Denmark}
\affiliation{DTU-Space, Technical University of Denmark, Elektrovej 327, DK-2800 Kgs. Lyngby, Denmark}

\author[0000-0002-4085-9165]{Carlos G{\'o}mez-Guijarro}
\affiliation{Universit{\'e} Paris-Saclay, Universit{\'e} Paris Cit{\'e}, CEA, CNRS, AIM, 91191, Gif-sur-Yvette, France}

\author{Katriona~M.~L.~Gould}
\affiliation{Cosmic Dawn Center (DAWN), Denmark}
\affiliation{Niels Bohr Institute, University of Copenhagen, Jagtvej 128, DK-2200 Copenhagen N, Denmark}

\author{Kasper E. Heintz}
\affiliation{Cosmic Dawn Center (DAWN), Denmark}
\affiliation{Niels Bohr Institute, University of Copenhagen, Jagtvej 128, DK-2200 Copenhagen N, Denmark}

\author{Olivier Ilbert}
\affiliation{Aix Marseille Univ, CNRS, CNES, LAM, Marseille, France}

\author[0000-0002-8896-6496]{Christian Kragh Jespersen}
\affiliation{Department of Astrophysical Sciences, Princeton University, Princeton, NJ 08544, USA}

\author{Vasily Kokorev}
\affiliation{Kapteyn Astronomical Institute, University of Groningen, P.O. Box 800, 9700AV Groningen, The Netherlands}
\affiliation{Cosmic Dawn Center (DAWN), Denmark}

\author{Mariko Kubo}
\affiliation{Astronomical Institute, Tohoku University, Aoba-ku, Sendai 980-8578, Japan}

\author{Georgios E. Magdis}
\affiliation{Cosmic Dawn Center (DAWN), Denmark}
\affiliation{DTU-Space, Technical University of Denmark, Elektrovej 327, DK-2800 Kgs. Lyngby, Denmark}
\affiliation{Niels Bohr Institute, University of Copenhagen, Jagtvej 128, DK-2200 Copenhagen N, Denmark}

\author{Conor McPartland}
\affiliation{Cosmic Dawn Center (DAWN), Denmark}
\affiliation{Niels Bohr Institute, University of Copenhagen, Jagtvej 128, DK-2200 Copenhagen N, Denmark}

\author{Masato Onodera}
\affiliation{Department of Astronomical Science, The Graduate University for Advanced Studies, SOKENDAI, 2-21-1 Osawa, Mitaka, Tokyo, 181-8588, Japan}
\affiliation{Subaru Telescope, National Astronomical Observatory of Japan, National Institutes of Natural Sciences (NINS), 650 North A’ohoku Place, Hilo, HI 96720, USA}

\author{Francesca Rizzo}
\affiliation{Cosmic Dawn Center (DAWN), Denmark}
\affiliation{Niels Bohr Institute, University of Copenhagen, Jagtvej 128, DK-2200 Copenhagen N, Denmark}

\author{Masayuki~Tanaka}
\affiliation{National Astronomical Observatory of Japan, 2-21-1 Osawa, Mitaka, Tokyo, 181-8588, Japan}
\affiliation{Department of Astronomical Science, The Graduate University for Advanced Studies, SOKENDAI, 2-21-1 Osawa, Mitaka, Tokyo, 181-8588, Japan}

\author{Sune Toft}
\affiliation{Cosmic Dawn Center (DAWN), Denmark}
\affiliation{Niels Bohr Institute, University of Copenhagen, Jagtvej 128, DK-2200 Copenhagen N, Denmark}

\author{Aswin P. Vijayan}
\affiliation{Cosmic Dawn Center (DAWN), Denmark}
\affiliation{DTU-Space, Technical University of Denmark, Elektrovej 327, DK-2800 Kgs. Lyngby, Denmark}

\author[0000-0003-1614-196X]{John R. Weaver}
\affiliation{Department of Astronomy, University of Massachusetts, Amherst, MA 01003, USA}

\author{Katherine E. Whitaker}
\affiliation{Department of Astronomy, University of Massachusetts, Amherst, MA 01003, USA}
\affiliation{Cosmic Dawn Center (DAWN), Denmark}

\author[0000-0002-3722-322X]{Lillian Wright}
\affiliation{Department of Physics and Astronomy, Seoul National University, 1 Gwanak-ro, Gwanak-gu, Seoul, South Korea}
\affiliation{Department of Astronomy, University of Massachusetts, Amherst, MA 01003, USA}

\begin{abstract}
We present the results of a systematic study of the rest-frame optical morphology of quiescent galaxies at $z \geq 3$ using the Near-Infrared Camera (NIRCam) onboard $JWST$. Based on a sample selected by $UVJ$ color or $NUVUVJ$ color, we focus on 26 quiescent galaxies with $9.8<\log{(M_\star/M_\odot)}<11.4$ at $2.8<z_{\rm phot}<4.6$ with publicly available $JWST$ data. Their sizes are constrained by fitting the S\'ersic profile to all available NIRCam images. We see a negative correlation between the observed wavelength and the size and derive their size at the rest-frame $0.5\, {\rm \mu m}$ using size measurements in multiple bands. Our quiescent galaxies show a significant correlation between the rest-frame $0.5\, {\rm \mu m}$ size and the stellar mass at $z\geq3$. The analytical fit for them at $\log{(M_\star/M_\odot)}>10.3$ implies that our size - stellar mass relations are below those at lower redshifts, with the amplitude of $\sim0.6\, {\rm kpc}$ at $M_\star = 5\times 10^{10}\, M_\odot$. This value agrees with the extrapolation of the size evolution of quiescent galaxies at $z<3$ in the literature, implying that the size of quiescent galaxies increases monotonically from $z\sim3-5$. Our sample mainly comprises galaxies with bulge-like structures according to their median S\'ersic index and axis ratio of $n\sim3-4$ and $q\sim0.6-0.8$, respectively. On the other hand, there is a trend of increasing fraction of galaxies with low S\'ersic index at higher redshift, suggesting $3<z<5$ might be the epoch of onset of morphological transformation with a fraction of very notable disky quenched galaxies.
\end{abstract}


\section{Introduction} \label{sec:intro}
\par One of the most fundamental ways to characterize galaxies is to look at their appearance. The size and morphology correlate with the formation history of galaxies, including major or minor mergers \citep[e.g.,][]{2009ApJ...697.1290B}, and their host halos. Thus, tracing them across cosmic time provides us with insights into the evolution of galaxies \citep[e.g.,][and references therein]{2014ApJ...788...28V}. 
\par The relationship between the galaxy half-light size in the rest-frame optical and stellar mass (hereafter called size-mass relation) has been examined up to $z\sim3$ \citep[e.g.,][]{2003MNRAS.343..978S,2014ApJ...788...28V,2017ApJ...839...71F,2019ApJ...880...57M,2021ApJ...921...38K,2022ApJ...925...34C}, showing that the size is larger for higher stellar mass galaxies. The slope of the size-mass relation is steeper for quiescent galaxies than that for star-forming galaxies, which could be due to the dry minor mergers \citep[see the discussion in e.g.,][]{2015ApJ...813...23V}. Moreover, these size-mass relations suggest that the size of quiescent galaxies drastically evolves as a function of redshift. For example, \citet{2014ApJ...788...28V} show that the size of quiescent galaxies at fixed stellar mass evolves as $\propto (1+z)^{-1.48}$ at $z<3$, whereas that of star-forming galaxies evolves as $\propto (1+z)^{-0.75}$. Various phenomena can cause this redshift evolution of size. Firstly, newly quenched galaxies at lower redshift having a larger size similar to star-forming galaxies can cause an apparent size evolution, which is called progenitor bias \citep[e.g.,][]{2013ApJ...775..106C, 2013ApJ...773..112C, 2013ApJ...765..104B}. The dry minor mergers also effectively increase the size of quiescent galaxies \citep[e.g.,][]{2012ApJ...744...63O,2012MNRAS.422.1714N}. In addition, this trend might be only in the half-light size. \citet{2019ApJ...877..103S} shows that the half-mass size does not show as strong redshift evolution compared to the half-light size mentioned above \citep[see also][]{2017ApJ...837....2M}. This implies that the observed light profile is different depending on the wavelength. To minimize its effect, some studies derive the half-light radius in the same rest-frame wavelength in comparing the different redshifts by using images spanning a range of observed-frame wavelengths \citep[e.g.,][]{2021ApJ...921...38K,2021MNRAS.506..928N}. 
\par Some quiescent galaxies at $z\sim2-3$ are reported as disk galaxies or spiral galaxies \citep[e.g.,][]{2017Natur.546..510T,2018ApJ...862..126N,2022ApJ...938L..24F}. In contrast, the majority of quiescent galaxies have a high S\'ersic index and axis ratio even at high redshift, which implies that they have bulge-like shapes \citep[e.g.,][]{2015ApJ...808L..29S, 2019ApJ...871..201M, 2021MNRAS.501.2659L}. Thus, there could be growing morphological variation for quiescent galaxies at high redshift.
\par Recently, massive quiescent galaxies have been found even at $z\gtrsim 3-4$ \citep{2017Natur.544...71G,2018A&A...618A..85S,2019ApJ...885L..34T,2020ApJ...903...47F,2020ApJ...889...93V,2021A&A...653A..32D,2022arXiv221211638N,2023MNRAS.520.3974C,2023arXiv230111413C}. The morphology of some of these galaxies are also explored using the ground-based telescopes. For example, \citet{2018ApJ...867....1K} stack the $K$-band images assisted by Adaptive Optics (AO) and derive the average rest-frame optical size of quiescent galaxies at $z\sim4$, which includes a spectroscopically confirmed quiescent galaxy at $z=4.01$ reported in \citet{2019ApJ...885L..34T} and \citet{2020ApJ...889...93V}. Also, the \textit{Hubble Space Telescope} ($HST$)/WFC3 images constrain the morphological properties of some $z\geq3$ spectroscopically confirmed quiescent galaxies \citep{2012ApJ...759L..44G,2020ApJ...905...40S, 2021MNRAS.501.2659L,2021ApJ...908L..35E}.
\par However, a statistical discussion of the structural properties was not carried out for quiescent galaxies at $z\geq3$, including the constraint on their size-mass relation and its redshift evolution to $z\sim0$. There are four underlying reasons. One is the requirement for high-resolution images which resolve these galaxies at $\geq2\, {\rm \mu m}$, corresponding to the rest-frame wavelength of $0.5\, {\rm \mu m}$ at $z\geq3$ often used in the size-mass relation at lower redshift. This wavelength range is not observable with $HST$. For $z>3$ galaxies, $HST$ can only probe the wavelength of $<0.4\, {\rm \mu m}$ in the rest frame. Other approaches, such as using AO, lead to expensive observation and a severely limited and incomplete sample. Secondly, since the method for selecting quiescent galaxies differs among the literature above \citep[e.g., based on the specific star formation rate or colors, also see Table 4 in][for the summary of the literature]{2023ApJ...947...20V}, comparisons between different samples are not straightforward. Thirdly, assembling a large sample of this rare galaxy population requires large and deep imaging, especially in near-infrared. Finally, constraining the wavelength dependence of the size of individual objects had been difficult. If there is a correlation between the structural properties and the wavelength for quiescent galaxies at $z\geq3$, as at low redshift, the measurement with a single identical band could cause a systematic bias due to the different redshift.
\par The \textit{James Webb Space Telescope} ($JWST$) is poised to revolutionize this topic. $JWST$'s high-resolution imaging at near-infrared, particularly at $\lambda>2\ \mu$m, allows us to probe the emission from the stellar populations of galaxies at $z\geq3$ and accurately measure their structural properties in the rest-frame optical. Moreover, utilizing its multi-band information across a wide range of wavelengths, we can evaluate the wavelength dependence of the sizes, as done at lower redshift. Indeed, several studies derive structural properties of distant quiescent galaxies with the Near-Infrared Camera (NIRCam) on $JWST$, such as for a quiescent galaxy at $z=4.658$ \citep{2023arXiv230111413C}, and a statistical sample of galaxies at to $z \sim 8$, including quiescent ones \citep{2022ApJ...937L..33S, 2024MNRAS.527.6110O}. The size evolution is also explored for star-forming galaxies at higher redshift \citep[e.g.,][]{2022arXiv220813582O,2022ApJ...938L..17Y, 2023arXiv230408517G,2023arXiv231102162W}.
\par In this paper, we systematically investigate the structural properties of a sample of 26 quiescent galaxies at $z\geq3$ from \citet{2023ApJ...947...20V} (hereafter denoted as V23) by using the $JWST$/NIRCam images. V23 select quiescent galaxies in fields with publicly available $JWST$ data based on $U-V$ and $V-J$ color \citep{2009ApJ...691.1879W} or $NUV - U$, $U - V$, and $V - J$ color \citep{2023arXiv230210934G}. This paper derives the rest-frame optical size-stellar mass relation and quantifies their typical morphological properties, focusing on the S\'ersic index and axis ratio. This paper is organized as follows. The sample selection for this study is summarized in Section \ref{sec:sample}. In Section \ref{sec:fit}, we summarize the method of morphological fitting and deriving the rest-optical size. Section \ref{sec:size} reports the size-mass relations of our quiescent galaxies and the analytical fit to them. The morphological properties, i.e., S\'ersic index and axis ratio, are discussed in Section \ref{sec:morph}. Lastly, we discuss and summarize our results in Section \ref{sec:dissum}. We assume the $\Lambda$CDM cosmology with $H_0=70\ {\rm km\ s^{-1}Mpc^{-1}}$, $\Omega_m = 0.3$, and $\Omega_\Lambda = 0.7$. The magnitude is based on the AB magnitude system \citep{1983ApJ...266..713O}. We assume the initial mass function of \citet{2003PASP..115..763C}.

\section{Sample Selection} \label{sec:sample}
\par This study uses the sample of quiescent galaxies at $z\geq3$ selected from $JWST$ data, constructed in V23. Here, we briefly overview the sample selection and refer the reader to V23 for more details.
\par V23 process the all publicly available $JWST$ imaging of fields spanning $\sim145\,  {\rm arcmin^2}$ taken during the first three months and extract sources therein using \textsc{Source Extractor Python (sep)} \citep{2016JOSS....1...58B}. Their photometric redshift, rest-frame colors, and stellar mass are estimated by \textsc{eazy-py} \citep{2008ApJ...686.1503B, Brammer_eazy-py_2021} with $JWST$ and $HST$ photometry. We confirm their photometric redshift agrees well with the spectroscopic redshift determined from the archival observations (G. Brammer et al., in preparation). In the CEERS field, we achieve a $\sigma_{NMAD}=0.0268$ in $\delta z/(1+z)$ for the spectroscopic sample. We select galaxies at $z\geq3$ from their photometric redshift. 
\par Two color selections for quiescent galaxies are employed. One is based on the traditional $UVJ$ diagram. V23 select galaxies located in the box of quiescent galaxies defined in \citet{2009ApJ...691.1879W}. We here do not consider the uncertainty of the color. This selection corresponds to the ``standard" selection argued in V23. The other is based on the $NUV - U$, $U - V$, and $V - J$ color diagram, hereafter called the $NUVUVJ$ diagram. This is recently proposed by \citet{2023arXiv230210934G}, which constructs a Gaussian Mixture Model for the quiescent galaxy selection at high redshift \citep{Gould_GMM-quiescent_2023}. Based on the $NUVUVJ$ diagram, this model returns the ``probability of being quiescent", $P_{\rm Q}$. \citet{2023arXiv230210934G} shows that the selection based on $P_{\rm Q}$ not only can reduce the contamination compared to the $UVJ$ diagram but also can select recently quenched galaxies more efficiently, which can be abundant at $z>3$. This study selects quiescent galaxies based on the $P_{\rm Q, 50\%}>0.7$, where $P_{\rm Q, 50\%}$ is the median of $P_Q$ derived from bootstrapping. We note that these selections of the $UVJ$ and $NUVUVJ$ colors are the most conservative ones among those in V23 not to be affected by the contaminants due to the uncertainty in color. These $UVJ$ and $NUVUVJ$ color selections give 34 and 18 quiescent galaxies, respectively. 16 quiescent galaxies are selected from both $UVJ$ and $NUVUVJ$ selections, which is likely because the threshold of $P_Q$ in $NUVUVJ$ selection is strict, and we select a similar population to those selected from $UVJ$.
\par We impose three additional selections to the sample of V23 to derive morphological properties accurately. Firstly, we require galaxies to be detected in more than three bands of $JWST$ with the signal-to-noise ratio of $S/N>50$ in $0.5\arcsec$ diameter aperture photometry. This criterion is motivated by a simulation using mock S\'{e}rsic profiles summarized in Section \ref{sec:mock} and ensures that we have reliable and precise structural parameter measurements (i.e., effective radius, S\'ersic index) in multiple bands. As described below, loosening this $S/N$ criterion will include more systematically biased output values. We here retain 30 and 18 galaxies selected by $UVJ$ or $NUVUVJ$, respectively. Second, we remove X-ray detections to avoid the contribution from active galactic nuclei (AGNs) to their images, which results in removing a galaxy (\#2718 in V23 ID) detected in the AEGIS-X Deep survey \citep{2015ApJS..220...10N} and a galaxy (\#2552 in V23 ID) detected in the Chandra observation of \citet{2014ApJS..212....9T}. They are selected only by $NUVUVJ$, leading to 30 and 16 galaxies selected by $UVJ$ or $NUVUVJ$, respectively. We note that none of the objects in our sample were listed in existing radio catalogs. Lastly, we do not use two quiescent galaxies from the SMACS0723 field to avoid overestimating the sizes due to the strong gravitational lensing. 
\par
\begin{figure*}
    \centering
    \includegraphics[width=16cm]{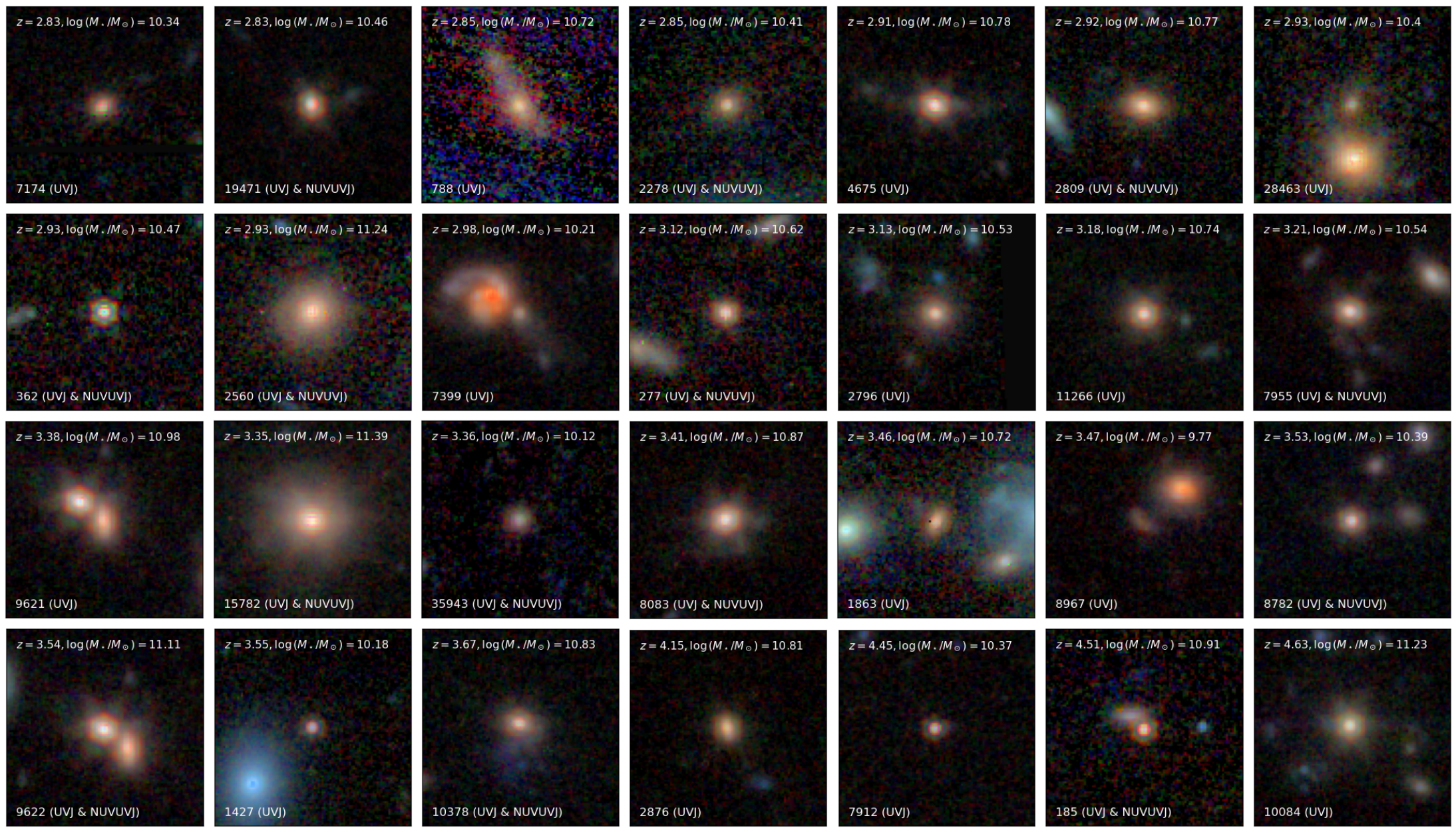}
    \caption{Color images of all quiescent galaxies used in this study. F277W, F356W, and F444W images are used for blue, green, and red, respectively. All images are $4.0\arcsec\times4.0\arcsec$ centered on each quiescent galaxy. They are sorted according to their photometric redshift. At the bottom of each figure, the ID of galaxies, identical to those of V23, and the method that selects it as quiescent one are shown. We can see that most of them have compact profiles.}
    \label{fig:img}
\end{figure*}

\begin{figure*}
    \centering
    \includegraphics[width=14cm]{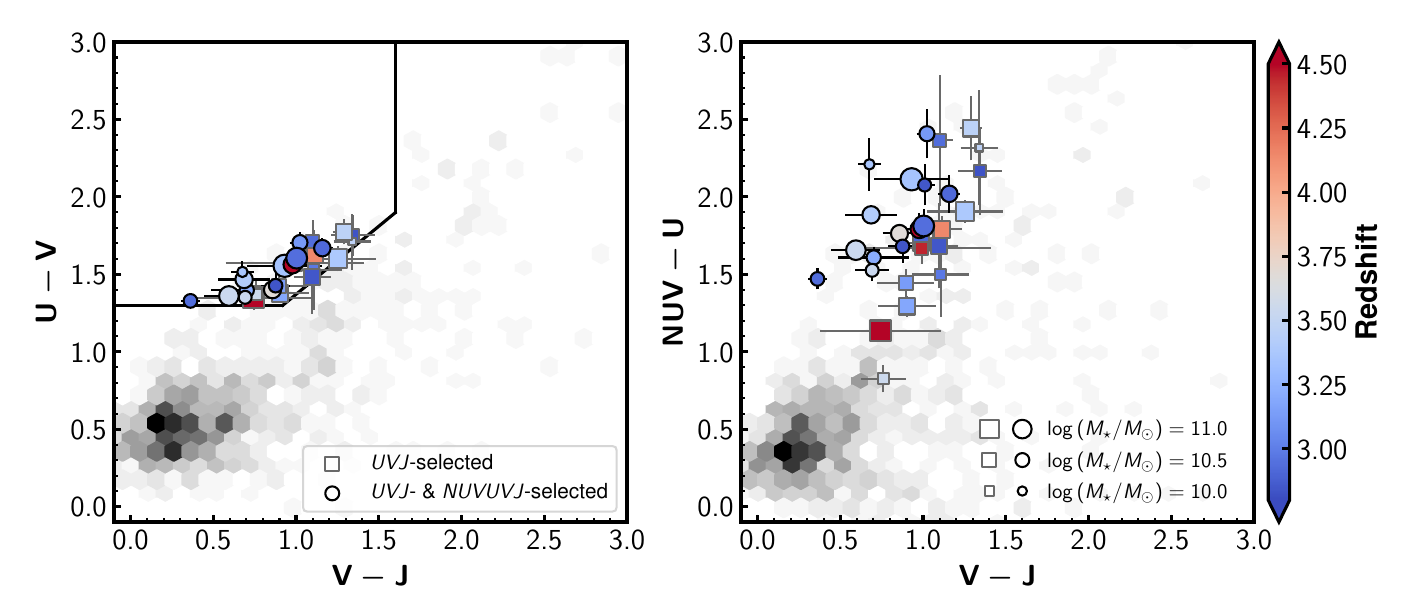}
    \caption{Left panel: $UVJ$ diagram \citep{2009ApJ...691.1879W}. Squares represent quiescent galaxies selected by the $UVJ$ only, whereas circles represent those selected by $UVJ$ and $NUVUVJ$. They are colored according to their photometric redshift, and the markers are larger for more massive galaxies. The gray background represents the distribution of galaxies with $\log{(M_\star/M_\odot)}>9.5$ at $3<z<5$. Right panel: $NUVUVJ$ diagram \citep{2023arXiv230210934G}. The meanings of the markers are the same as in the left panel.}
    \label{fig:colordiagram}
\end{figure*}

 In total, we use 28 quiescent galaxies selected by at least either $UVJ$ selection or $NUVUVJ$ selection. All of these galaxies are selected by $UVJ$, and 14 galaxies are selected by $NUVUVJ$. They are at $2.83\leq z \leq 4.63$ with $9.77\leq \log{(M_\star/M_\odot)}\leq11.39$ according to \textsc{eazy-py}. Figure \ref{fig:img} summarizes the RGB images of the long wavelength of NIRCam for these galaxies. Their location in the $UVJ$ and $NUVUVJ$ diagram are shown in Figure \ref{fig:colordiagram}. We refer the reader to the supplement data of V23 for their SED \citep{francesco_valentino_2023_7614908}. The sample includes galaxies in CEERS \citep[ERS \#1345, PI: S. Finkelstein,][]{2023ApJ...946L..12B}, Stephan's Quintet \citep[ERO \#2736,][]{2022ApJ...936L..14P}, PRIMER (PRIMER, GO \#1837, PI: J. S. Dunlop), North Ecliptic Pole Time-Domain Field (GTO \#2738, PI: R. Windhorst), J1235 (COM/NIRCam \#1063, PI: B.Sunnquiest), and SGAS1723 (as a part of TEMPLATES, ERS \#1355, PI: J. Rigby).

\section{Morphological Fitting} \label{sec:fit}
\par To obtain the structural parameters, we fit an analytical function to the $JWST$/NIRCam images. Though there could be asymmetrical features or two components, such as bulge and disk, we simply fit the single S\'ersic profile \citep{1963BAAA....6...41S} following the literature \citep[e.g.,][and others]{2014ApJ...788...28V,2019ApJ...880...57M, 2021MNRAS.506..928N} to fairly compare them. We note that the reduced chi-square of the fitting described in Section \ref{sec:3.3} implies that this assumption is reasonable.
\subsection{Input Images} \label{sec:img}
\par We use all available images of $JWST$/NIRCam from the F115W filter to the F444W filter (see Table 3 of V23 for the full list of available filters on each field). We use each filter's cutouts of $4.0\arcsec\times4.0\arcsec$ as input object images. The variance images generated by the \textsc{Grizli} pipeline \citep{2018zndo...1146904B,2022zndo...6672538B} are used to make sigma images. The sigma image can sometimes be overestimated or underestimated; thus, it is renormalized by a correction factor, which makes the sigma image of the regions without any sources equal to its fluctuation. Object masks are generated using the {\tt photutils} module. Objects more than $1\arcsec$ away from the target objects are masked.
\subsection{Point Spread Function} \label{sec:PSF}
\par The point spread function (PSF) is an essential factor in conducting the morphological fitting precisely. Previous studies use the theoretical PSFs \citep[e.g.,][]{2022ApJ...937L..33S} based on the {\tt WebbPSF} software \citep[][]{2012SPIE.8442E..3DP,2014SPIE.9143E..3XP}, those based on the software with the real images \citep[e.g.,][]{2023arXiv230413776Z}, such as \textsc{PSFEx} \citep{2011ASPC..442..435B}, or natural stars \citep[e.g.,][]{2022ApJ...939L..28D,2022ApJ...938L..17Y}. This study utilizes the theoretical PSFs from {\tt WebbPSF} smoothed to match their radial flux profile to the median profile of natural stars of each field. This approach copes with the possible difference in the profile between the theoretical PSFs and the natural stars reported recently \citep{2022ApJ...939L..28D,2022arXiv220813582O} and is free from the noises that exist in the stacked star images.
\par The theoretical PSFs are first generated for each filter with {\tt WebbPSF}. Here, in-flight sensing measurements are used by retrieving the Optical Path Difference files at the nearest dates to the observation date. The obtained PSFs are rotated to match the position angle. Next, we derive the radial flux profile of natural stars. Stars are selected from the source catalog of each field by focusing on the unresolved objects based on the {\tt FLUX\_RADIUS} and with the small ellipticity from \textsc{sep}, which is a similar approach to that employed in \textsc{PSFEx}. The detailed selection is summarized in Appendix \ref{sec:app1}. We then smooth the PSFs from {\tt WebbPSF} with the Moffat profile to match its one-dimensional radial profile with that of the median-stacked natural stars. The half width at half maximum of the used PSF is $0.03\arcsec-0.08\arcsec$, which is larger at a longer wavelength (see Table \ref{tab:2}). Figure \ref{fig:PSF} shows the one-dimensional radial profiles of smoothed PSFs used in this study in CEERS as an example.
\subsection{Fitting} \label{sec:3.3}
\par In fitting the S\'ersic profile to images, we use \textsc{Galfit} \citep{2002AJ....124..266P,2010AJ....139.2097P}. It uses real object images, sigma images, mask images, and PSFs as inputs and returns the best-fit parameters. We first fit the F277W image of each galaxy, which shows the light profile at around the rest-frame $0.5\, {\rm \mu m}$, similar to \citet{2014ApJ...788...28V} for our galaxies ($\lambda_{\rm rest}=0.49-0.72\, {\rm \mu m}$ depending on their redshift). In the fitting, the magnitude $m$, effective radius $r_{\rm eff}$, S\'ersic index $n$, axis ratio $q$, position angle PA, and the central pixel coordinates are set as the free parameters. We allow these parameters in the following range to avoid the unrealistic cases: $-3\, {\rm mag}<m-m_{\rm SEP}<3\, {\rm mag}$, $0.1\, {\rm pixel}<r_{\rm eff}<500\, {\rm pixel}$, $0.2<n<8$, and $0.0001 <q< 1$, where $m_{\rm SEP}$ is the Kron magnitude from the \textsc{sep}. Next, we fit the images in the other filters. In this case, we fix the axis ratio and position angle as the best-fit value in the F277W images as in \citet{2021MNRAS.506..928N} since they are not likely to depend on the wavelength. This enables us to derive the effective radius and the S\'ersic index consistently in the different wavelengths. We get the S\'ersic index in the edge of the parameter range ($n=8$) for some galaxies. In that case, we refit the images of all filters, assuming $n=4$. The sky background of the images is also set to be a free parameter, but the result does not change significantly even if we fix it to the value derived independently. Also, surrounding objects within $1\arcsec$ from the target galaxies are simultaneously fit by the S\'ersic profile. 
\par
\begin{figure}[ht]
    \centering
    \includegraphics[width=8.5cm]{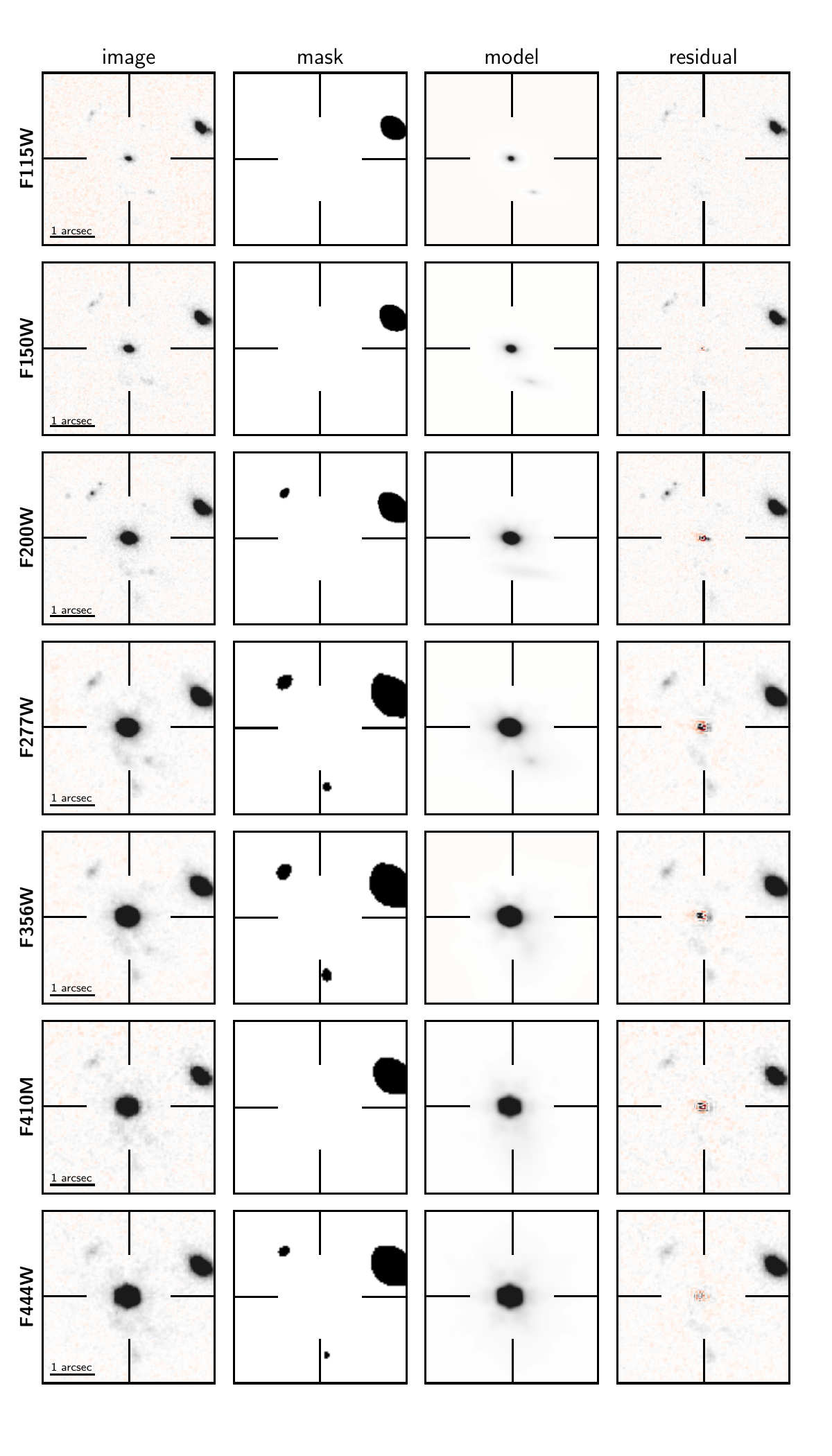}
    \caption{A \textsc{Galfit} fit of \#7995 as an example, which is $UVJ$-selected quiescent galaxy at $z=3.21$ with $\log{(M_\star/M_\odot)}=10.54$ in CEERS field. The observed object, mask, model, and residual images are shown from left to right panels in each row. The longer wavelength images are shown toward the bottom row. All images are $4.0\arcsec\times4.0\arcsec$. The images are scaled by the noise ($\pm15\sigma$) of object images in each filter. The apparent size in these images does not necessarily reflect the actual size of galaxies due to the image scaling and the PSF smoothing.}
    \label{fig:fit}
\end{figure}

 Figure \ref{fig:fit} shows an example of the result of the fitting for all available images of a galaxy, which is the typical one of our sample with the redshift and the stellar mass close to the median values. Appendix \ref{app:2} also summarizes the fitting results in the F200W filter and F277W filter and the best-fit parameters in the F277W filter of all galaxies used in Section \ref{sec:size} and \ref{sec:morph} (see Section \ref{sec:cg} for the selection). Some galaxies have compact residuals with both positive and negative pixels. These residuals are often seen in $HST$ images of compact galaxies with high S\'ersic index, too \citep[e.g.,][]{2019ApJ...871..201M, 2021MNRAS.501.2659L,2022ApJ...938..109F}. Most residuals are just 10\% of the flux of the original images at most. The median reduced chi-square of the fitting for targets is $\chi^2_{\nu}= 1.0-1.2$ in each filter, suggesting that the assumption of the single S\'ersic profile is overall good.
\par To estimate realistic uncertainties of the obtained parameters, we conduct the following procedure. The PSF-convolved S\'ersic profile of the best-fit parameters for each galaxy is first inserted into a noise image for each filter and each field. The noise images are created by randomly cutting out the real images without any sources. We apply \textsc{Galfit} to this mock image in the same manner as for the real image and obtain the best fit. This procedure is repeated for 100 different noise images, and the uncertainty of each parameter is taken as the range between the 16th and 84th values of this distribution. We find that the original uncertainty reported by \textsc{Galfit} is 2-5 times more underestimated depending on wavelength than that derived by this procedure. 
\subsection{Bias and quality in the fitting}\label{sec:mock}
\par The quality of the \textsc{Galfit} best-fit model depends on the signal-to-noise ratio of the images \citep[e.g., Figure 6 in ][]{2012ApJS..203...24V}. Thus, the different brightness and different profiles of galaxies will have a different bias and quality in the fitting. Here, we evaluate the bias and quality of the fitting using mock galaxy images following the S\'ersic profile.
\par Firstly, 3000 different single S\'ersic profiles with various ranges of parameters are prepared. Parameters for each profile are randomly selected in the following range: $21\, {\rm mag}<m<29\, {\rm mag}$, $0.002\arcsec<r_{\rm eff}<2\arcsec$, and $0.1 <q< 1$. The position angle is set to be random. Since we are focusing on quiescent galaxies, which are like to have bulge-like structures supported by the literature \citep[e.g.,][]{2021MNRAS.501.2659L} and Section \ref{sec:n} of this paper, this section assumes S\'ersic index as $n=3-5$. For reference, Appendix \ref{app:mockn13} shows the case of $n=1-3$. Its trend is broadly consistent with the case with $n=3-5$, except for the smallest input size ($0.002\arcsec<r_{\rm eff}<0.02\arcsec$). Next, after convolving with the corresponding PSF, we insert them into 30 different noise images for each. In this section, the CEERS F277W image is used as an example. We note that the results do not significantly change even if we use images in different filters. In total, we generate 90000 mock S\'ersic profile images. We fit the suite of mock galaxies using \textsc{Galfit} in the same manner as for the real galaxies and compare the obtained parameters with the input parameters.
\par
\begin{figure*}[t]
    \centering
    \includegraphics[width=17cm]{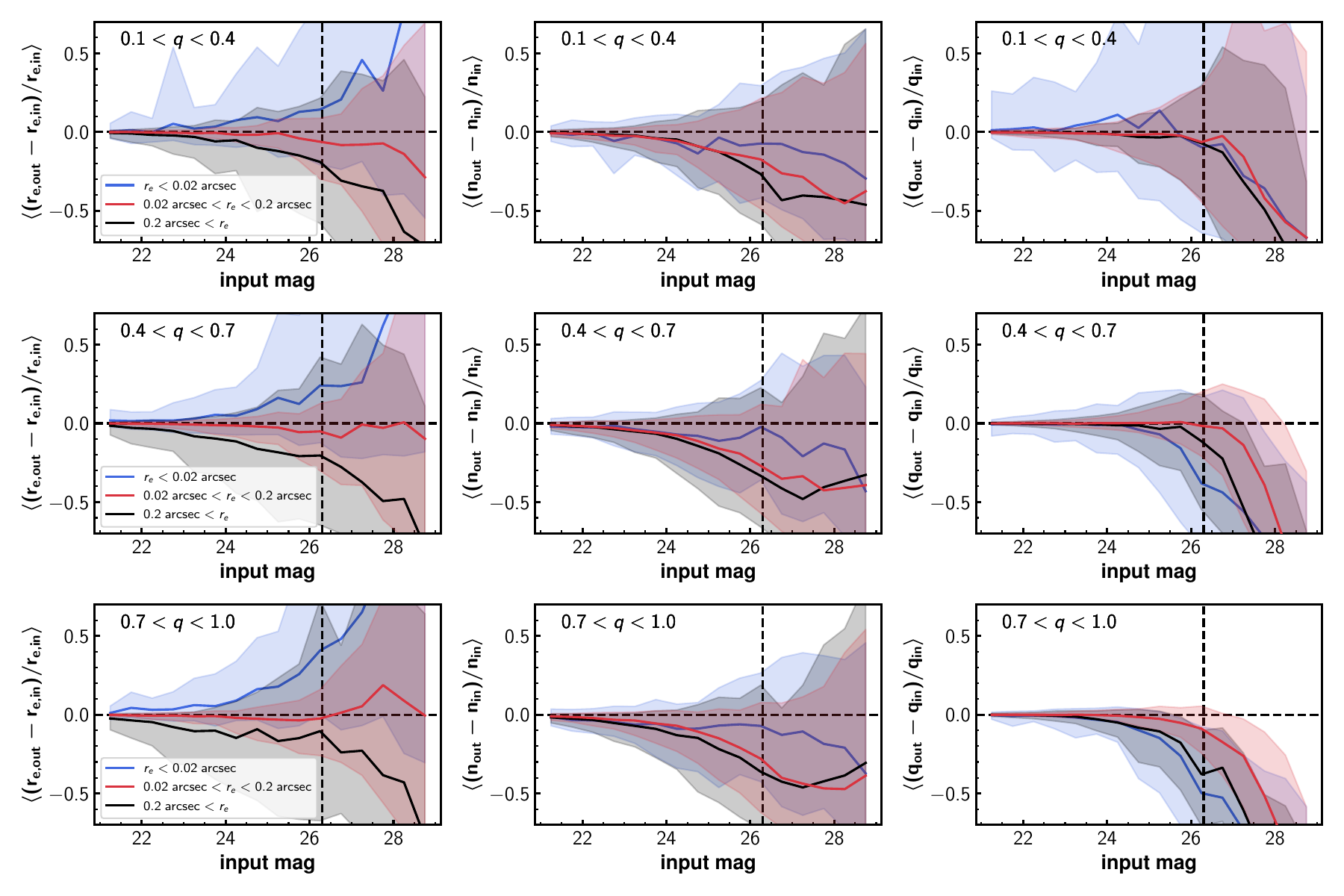}
    \caption{Median relative uncertainties of the estimated parameters (effective radius: left, S\'ersic index: middle, axis ratio: right) derived through mock galaxy images with S\'ersic index of $n=3-5$ as a function of the magnitude. The top, middle, and bottom panel corresponds to the case with the input axis ratio of $0.1<q<0.4$, $0.4<q<0.7$, and $0.7<q<1.0$, respectively. Different colors correspond to different input effective radii, i.e., $0.002\arcsec<r_{\rm eff}<0.02\arcsec$, $0.02\arcsec<r_{\rm eff}<0.2\arcsec$, and $0.2\arcsec<r_{\rm eff}<2\arcsec$. The shaded regions correspond to the ranges between the 16th and 84th values. The vertical line in each panel corresponds to the $S/N=50$ in the CEERS F277W image.}
    \label{fig:mock}
\end{figure*}

 Figure \ref{fig:mock} shows the relative uncertainty of the effective radius, S\'ersic index, and axis ratio. As an example, we here focus on the trend at $\sim26.3$ mag, which corresponds to the $S/N=50$ detection in the CEERS F277W image (see V23 for the depth of images). Firstly, the effective radii are slightly overestimated in smaller input sizes (20\% in $0.002\arcsec<r_{\rm eff}<0.02\arcsec$) and slightly underestimated in larger input sizes (20\% in $0.2\arcsec<r_{\rm eff}<2\arcsec$). The S\'ersic index is underestimated ($10-40\%$, equal to $\delta n\sim0.4-1.6$ for $n=3-5$) regardless of the input size and axis ratio. The axis ratio is well reproduced for $0.02\arcsec<r_{\rm eff}<0.2\arcsec$, but underestimated in larger or smaller sizes, in particular larger axis ratio ($q>0.4$). Fainter (brighter) objects are expected to have more (less) significant scatter and bias in all parameters.
\par As described in the following sections, our sample typically has $r_{\rm eff}\sim0.6$ kpc ($\sim 0.1\arcsec$) and $q\sim0.6-0.8$. Therefore, this test implies that our sample with our signal-to-noise ratio cut $S/N>50$ is typically expected to have only negligible offset in the effective radii and axis ratio ($<10\%$) and a slight offset in the S\'ersic index ($< 30\%$). 
\subsection{Wavelength dependence of the size, rest-frame optical sizes}\label{sec:cg}
\par It is known that the size of galaxies has a dependency on the observed wavelength at $z<3$ \citep[e.g.,][]{2014ApJ...788...28V,2021ApJ...921...38K,2022ApJ...937L..33S}. Our sample spans a wide redshift range; thus, obtaining the effective radius in the same rest-frame wavelength for all galaxies is important for a fair comparison with the literature at lower redshift. 
\par We obtain the effective radius in the rest-frame $0.5\, {\rm \mu m}$ and the wavelength dependence of the effective radius by fitting the measured effective radii in the different wavelengths to the following form: 
\begin{equation}
    \log{r_{\rm eff}(\lambda_{\rm rest})} = \gamma \log{\left(\frac{\lambda_{\rm rest}}{0.5\, {\rm \mu m}}\right)} + \log{r_{\rm eff,0.5\mu m}},
    \label{eq:wavre}
\end{equation}
where $\gamma$ and $r_{\rm eff,0.5\mu m}$ correspond to the strength of the wavelength dependence of the effective radius ($\gamma=\Delta\log{r_{\rm eff}}/\Delta \log{\lambda_{\rm rest}}$) and the effective radius at the rest-frame $0.5\, {\rm \mu m}$. $\lambda_{\rm rest}$ is the wavelength in the rest frame according to the photometric redshift. We note that the uncertainty in photometric redshift (see Section \ref{sec:sample}) induces only negligible impact on the rest-frame wavelength $\lambda_{\rm rest}$ ($\sim3\%$). Following the literature, we use the effective radius along the semi-major axis as the proxy of the galaxy size. In the fitting, we use the effective radii of the images in all bands where the target source is detected with $S/N>50$. This detection threshold is to assure the measurement quality in each band and motivated by the mock simulation described in Section \ref{sec:mock}. In addition, the best effective radii at the edge of the parameter range (i.e., $r_{\rm eff} = 0.1$ pixel or $500$ pixel) are not used here. After these selections, we require that at least three bands be used. The number of data points in this fitting depends on the field and the target's brightness, which spans from three to seven. 
\par We successfully determine $\gamma$ and $r_{\rm eff,0.5\mu m}$ of 26 quiescent galaxies, which we use in the following analysis. Two galaxies (\#7912 and \#362) are removed from the discussion because they are too compact and have the best effective radii at the edge of the parameter range in all wavelengths at least longer than $2.7\, \mu$m and thus do not have enough number of size measurements. Among 26 galaxies, all of them are selected by the $UVJ$ diagram, and 13 of them are selected by the $NUVUVJ$ diagram. We remind the reader that Appendix \ref{app:2} summarizes the fitting result and the best-fit parameters of all of them.
\par
\begin{figure*}
    \centering
    \includegraphics[width=16cm]{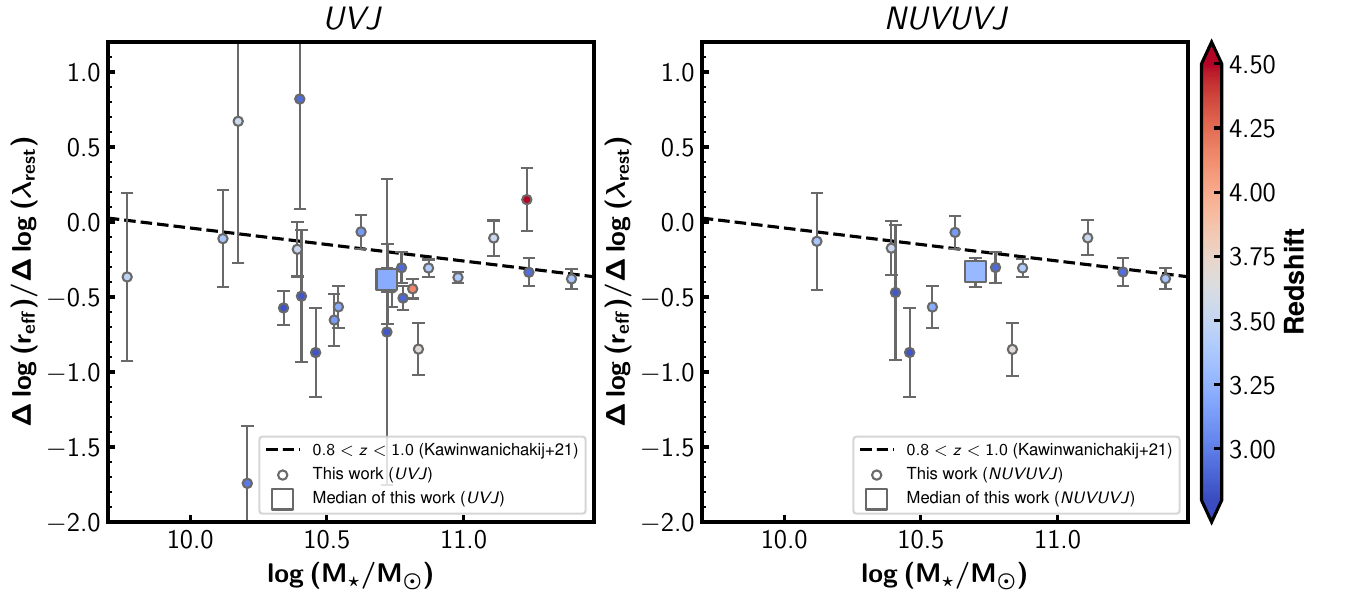}
    \caption{Wavelength dependence of the effective radius ($\gamma=\Delta\log{r_{\rm eff}}/\Delta \log{\lambda_{\rm rest}}$) as a function of stellar mass for $UVJ$ (left) and $NUVUVJ$ (right) selected quiescent galaxies. \#185 is not shown since its wavelength dependence is hardly constrained. Each marker is colored according to its redshift. Large squares show the median of each sample. A dashed line is the relation of quiescent galaxies at $0.8<z<1.0$ in \citet{2021ApJ...921...38K}.}
    \label{fig:wavegrad}
\end{figure*}

 Figure \ref{fig:wavegrad} shows the strength of the wavelength dependence of the effective radius $\gamma$ as a function of the stellar mass. Most quiescent galaxies have a negative correlation between the wavelength and the size, i.e., smaller sizes in longer wavelengths. The median values of $UVJ$-selected and $NUVUVJ$-selected sample is $\gamma = -0.38_{-0.08}^{+0.08}$ and $\gamma = -0.33_{-0.11}^{+0.10}$, respectively. This negative correlation between wavelength and the size agrees with the lower redshift results \citep[e.g.,][]{2014ApJ...788...28V,2021ApJ...921...38K,2022ApJ...937L..33S}. \citet{2014ApJ...788...28V} report the average strength of the wavelength dependence of the effective radius of quiescent galaxies at $0<z<2$ as $\gamma = -0.25$, less significant than our value. This could be due to the redshift evolution or the different stellar mass distribution between our sample and \citet{2014ApJ...788...28V}. Given the small size of our sample, we do not attempt to model the correlation between the stellar mass and the strength of the wavelength dependence of the effective radius as reported in \citet{2021ApJ...921...38K}, but we note that most of our sample is in good agreement with the relationship at the highest redshift bin ($0.8<z<1.0$) in \citet{2021ApJ...921...38K} with some outliers with the stronger negative gradient at lower redshift and lower mass.

\section{Rest-frame Optical Size-mass relation} \label{sec:size}
\subsection{Overall Trends}\label{sec:4p1}
\par
\begin{figure*}[t]
    \centering
    \includegraphics[width=16cm]{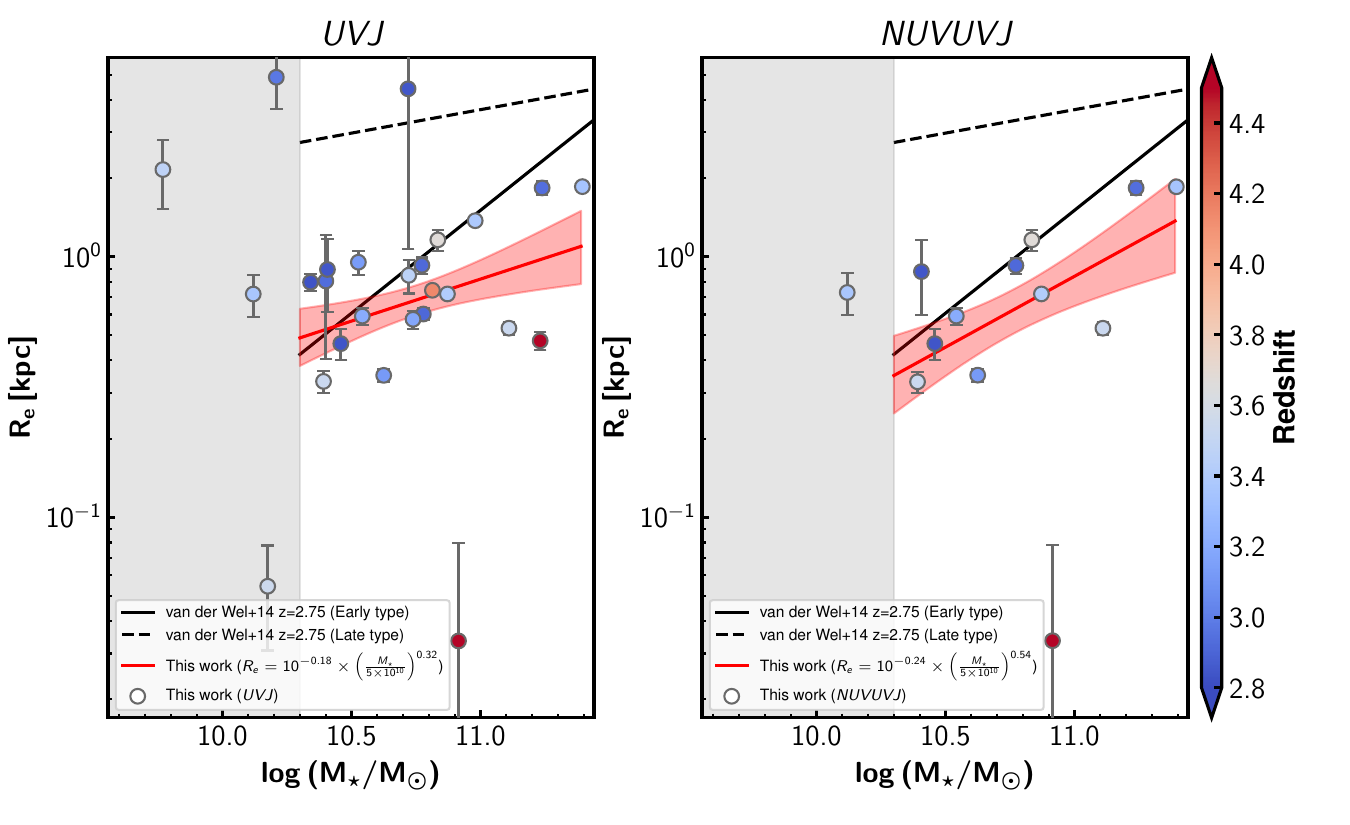}
    \caption{Size-mass relation of $UVJ$-selected (left) and $NUVUVJ$-selected (right) quiescent galaxies at $z\geq3$. The effective radii are measured at the rest-frame $0.5\,\mu$m. Circles represent quiescent galaxies from this study, color-coded by their photometric redshift. The red line and shaded region are the best fit of the size-mass relation of quiescent galaxies from this study and its $1\sigma$ uncertainty, respectively. The data points located in the gray hatched region ($\log{(M_\star/M_\odot)}<10.3$) are not used for the analytical fit. The solid and dashed black line is the best fit of the size-mass relation of early-type (i.e., quiescent) and late-type (i.e., star-forming) galaxies at $2.5<z<3.0$ in \citet{2014ApJ...788...28V}, respectively.}
    \label{fig:sizemass}
\end{figure*}

 We show quiescent galaxies at $z\geq3$ in the size-mass plane in Figure \ref{fig:sizemass}. The effective radii at the rest-frame $0.5\, {\rm \mu m}$ obtained in Section \ref{sec:cg} are used. First of all, we can see that many quiescent galaxies of our sample have effective radii of less than $1$ kpc, which is significantly smaller than those of star-forming galaxies at $z<3$ \citep[e.g.,][]{2014ApJ...788...28V}. Their effective radii appear to correlate with their stellar mass, i.e., more massive galaxies have larger radii. This is consistent with the size-mass relation of quiescent galaxies in lower redshift studies \citep[e.g.,][]{2014ApJ...788...28V,2019ApJ...880...57M}, which will be discussed in the following section. 
\par We see that several galaxies deviate from the trend. Some of the less massive galaxies ($\log{(M_\star/M_\odot)}<10.5$) have $r_{\rm eff}>2\, {\rm kpc}$ in the $UVJ$-selected case (\#788, \#7399, \#8967). This may mean that these galaxies are in a period of morphological transformation from star-forming galaxies and larger than quiescent galaxies or the contaminants of star-forming galaxies. On the other hand, we can not rule out that this is due to the difficulty in fitting the S\'ersic profile to these galaxies. For example, \#788 has an elongated structure, which can be a sign of a merger (Figure \ref{fig:fit-summary_lowz}). Also, \#7399 can be affected by nearby sources. \#8967 has an arclet shape, but it is not likely strongly lensed since the redshift of the companion galaxy ($z=3.49$) is consistent with \# 8967. We note that removing these galaxies from the sample does not affect our conclusions. In contrast, there are galaxies with significantly small sizes ($r_{\rm eff}<0.1\, {\rm kpc}$, \#185 and \#1427 in PRIMER). AGNs can cause the apparent compact morphology, but we do not have clear evidence that this is due to their contribution because they are not detected in X-ray \citep[X-UDS,][]{2018ApJS..236...48K}. The limiting flux of X-UDS survey is $1.4\times10^{-16}\, {\rm erg/cm^2/s}$ in the $0.5-2\, {\rm keV}$. This corresponds to the limiting luminosity in the rest-frame $2-10\, {\rm keV}$ of $L_X = (1.5-1.6)\times10^{43}\, {\rm erg/s}$ at their redshift, which implies that they are at least not likely powerful unobscured X-ray AGNs. We find that removing these galaxies from the sample does not affect our conclusions, either.
\par  We note that our stellar mass is robust against the codes we use. The stellar mass of our sample measured with \textsc{eazy-py} is compared to that measured with \textsc{FAST++}\footnote{\url{https://github.com/cschreib/fastpp}}. They are in good agreement with the offset of $\log{(M_{\rm \star, EAZY}/M_\odot)}-\log{(M_{\rm \star, FAST++}/M_\odot)} = 0.069_{-0.066}^{+0.075}\ (0.067_{-0.027}^{+0.033})$ for our $UVJ$ ($NUVUVJ$)-selected quiescent galaxies, respectively. \citet{2014ApJ...788...28V}, \citet{2019ApJ...880...57M} and \citet{2021MNRAS.506..928N}, whose size-mass relation will be compared with ours, use the stellar mass measured with \textsc{FAST}. Thus, this supports that there is no significant bias in comparing our size-mass relation with them.
\subsection{Analytic fits} \label{sec:sizemass-fit}
\par Motivated by the appearance of the existence of size-mass relation, we attempt to analytically fit our sample following the literature \citep[][]{2014ApJ...788...28V, 2019ApJ...880...57M, 2021MNRAS.506..928N, 2021ApJ...921...38K}. Firstly, following \citet{2014ApJ...788...28V}, the effective radii distribution is assumed to be a log-normal distribution $N(\log{R_{\rm eff}},\sigma_{\log{R_{\rm eff}}})$, where $\log{R_{\rm eff}}$ is the mean of the effective radius and $\sigma_{\log{R_{\rm eff}}}$ is the intrinsic scatter of the distribution. We assume that the mean of the effective radius $R_{\rm eff}$ depends on the stellar mass as:
\begin{equation}
    R_{\rm eff}/{\rm kpc} = A (\frac{M_\star}{5\times10^{10}M_\odot})^\alpha,
    \label{eq:sizemass}
\end{equation}
where $A$ is the effective radius at $M_\star = 5\times10^{10}\, M_\odot$ and $\alpha$ is the slope of the size-mass relation. We consider the uncertainty $\delta r_{\rm eff}$ of the observed effective radius $r_{\rm eff}$ by assuming it follows a Gaussian distribution. The $\delta r_{\rm eff}$ includes not only the observed uncertainty of $r_{\rm eff}$ but also the uncertainty of the stellar mass, assuming $0.21$ dex, the scatter of stellar mass from \textsc{eazy-py} \citep[][and also see the discussion in Section \ref{sec:4p1}]{2023arXiv230210934G}. The latter is included by converting it to the uncertainty of the effective radius by assuming the typical slope ($\alpha=0.6$) of the size-mass relation at lower redshift \citep[e.g.,][]{2019ApJ...880...57M}.
\par Following \citet{2014ApJ...788...28V}, we calculate the total likelihood for three parameters ($A$, $\alpha$, and $\sigma_{\log{R_{\rm eff}}}$) as 
\begin{equation}
\mathcal{L} = \Sigma \ln \left[W \cdot \langle  N(\log{R_{\rm eff}},\sigma_{\log{R_{\rm eff}}}), N(\log{r_{\rm eff}},\delta \log{r_{\rm eff}}) \rangle\right],
\end{equation}
where $W$ is the weight to avoid the dominance of lower-mass galaxies. We use the inverse of the stellar mass function of quiescent galaxies at $3<z<3.5$ in \citet{2022arXiv221202512W} as the weight. The stellar mass function itself has uncertainty due to effects such as cosmic variance \citep[e.g.,][]{2011ApJ...731..113M,2021ApJ...923....8S}, SED modeling uncertainties \citep[e.g.,][]{2009ApJ...701.1765M,2023ApJ...944..141P}, photometric redshift, and Poisson noise. The weight fluctuates according to the uncertainty of the stellar mass function to include these effects.
\par Several studies report that the size-mass relation bends at $\log{(M_\star/M_\odot)}=10-10.5$ at $z<2$ \citep[e.g.,][]{2019ApJ...872L..13M, 2021MNRAS.506..928N, 2021ApJ...921...38K,  2022ApJ...925...34C}. To avoid the deviation from our assumption in Equation \ref{eq:sizemass}, we only fit the galaxies with $\log{(M_\star/M_\odot)}>10.3$. This is the same stellar mass limit imposed in \citet{2014ApJ...788...28V}. In addition, dusty star-forming galaxies could be contaminating our sample \citep[e.g.,][]{2023ApJ...946L..16P}. They could have larger rest-frame optical sizes than typical quiescent galaxies \citep[e.g., $\sim2\, {\rm kpc}$ at $z\sim3$ in][]{2023arXiv230317246G}, driving a larger scatter of the relation. Thus, we randomly remove 20\% of the galaxies from the sample in the fitting, which is the contamination fraction of $UVJ$ quiescent galaxies in \citet{2018A&A...618A..85S}. Following the literature \citep[e.g.,][]{2014ApJ...788...28V}, we randomly select those to be removed by weighting galaxies by the ratio of the stellar mass function of star-forming galaxies and that of quiescent galaxies at $3<z<3.5$ in \citet{2022arXiv221202512W}. Even if we do not account for these potential contaminants, we find that the best-fit parameters are consistent within their $1\sigma$ uncertainty. In the fitting, the Markov Chain Monte Carlo (MCMC) method is employed using {\tt emcee} \citep{2013PASP..125..306F}. The fitting procedure, including fluctuating the weight $W$ and randomly removing the possible contaminants, is repeated 500 times. The best-fit values of A (the effective radius at $M_\star=5\times10^{10}M_\odot$), $\alpha$ (the slope of the size-mass relation), and $\sigma_{\log{R_{\rm eff}}}$ (the intrinsic scatter of the size-mass relation) are obtained from the total of their chain.
\par Figure \ref{fig:sizemass} shows the best-fit of the size-mass relation, and Table \ref{tab:1} summarizes the best-fit values of the parameters. The best-fit $\alpha$ confirms that the positive correlation between the stellar mass and the effective radius exists in both $UVJ$-selected and $NUVUVJ$-selected galaxies, even considering the $1\sigma$ uncertainty. The intrinsic scatter is $\sim0.2$ dex for both the $UVJ$-selected and $NUVUVJ$-selected sample, similar to those reported at $z<3$ \citep[e.g.,][]{2014ApJ...788...28V,2021ApJ...921...38K}.
\par
\begin{figure*}
    \centering
    \includegraphics[width=16cm]{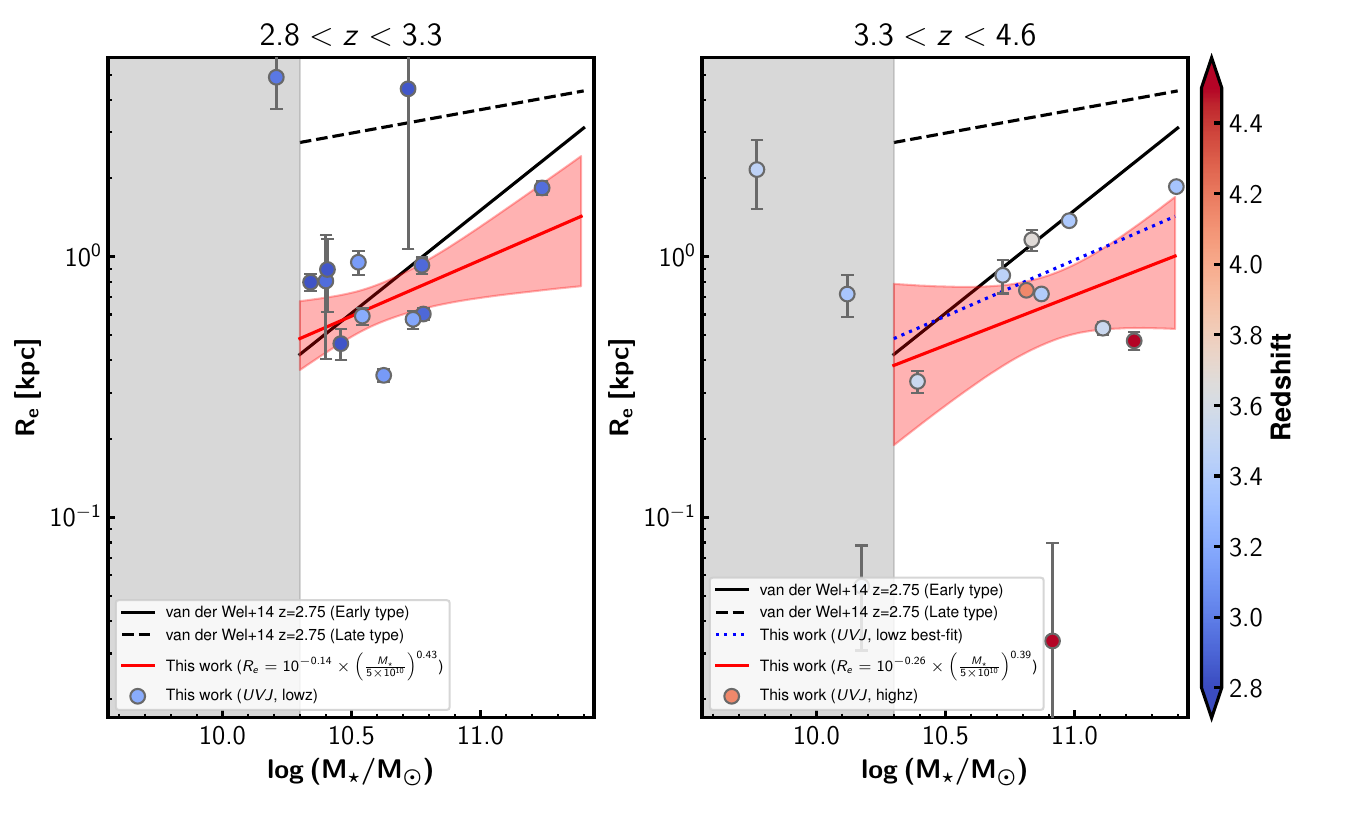}
    \caption{Size-mass relation of $UVJ$-selected quiescent galaxies split into two subsamples, $z<3.3$ (left) and $z>3.3$ (right). The meanings of the markers and the lines are the same as in Figure \ref{fig:sizemass}. In the right panel, the best-fit of $z<3.3$ subsample is shown in a dashed blue line for easy comparison.}
    \label{fig:sizemass-zbin}
\end{figure*}

 We split $UVJ$-selected quiescent galaxies into two subsamples based on their photometric redshift, i.e., $z<3.3$ and $z>3.3$. The threshold corresponds to the median of the redshift of the sample. Figure \ref{fig:sizemass-zbin} shows the size-mass relation of these subsamples and their best fit. Table \ref{tab:1} summarizes the best-fit values of the parameters. Though the uncertainty is more considerable than the total $UVJ$-selected sample, we see a positive correlation between the stellar mass and the effective radius even in these subsamples. The effective radius at the fixed stellar mass is smaller in the high-redshift sample than in the low-redshift sample. This suggests a potential redshift evolution of the sizes of quiescent galaxies at $z\geq3$.
\begin{deluxetable}{c|c||ccc}
\tabletypesize{\footnotesize}
\tablecaption{Summary of the obtained best-fit parameters of the size-mass relation.}\label{tab:1}
\tablehead{
\colhead{Sample} & \colhead{N\tablenotemark{a}} & \colhead{$\log{(A/{\rm kpc})}$\tablenotemark{b}} & \colhead{$\alpha$\tablenotemark{b}} & \colhead{$\sigma_{\log{R_{\rm eff}}}\tablenotemark{b}$}}
\startdata
     $UVJ$ & 22 & $-0.18_{-0.06}^{+0.06}$ &$0.32_{-0.21}^{+0.20}$ & $0.18_{-0.05}^{+0.07}$\\ 
     $UVJ$ ($z<3.3$) & 12 & $-0.14_{-0.08}^{+0.08}$ & $0.43_{-0.35}^{+0.29}$ & $0.16_{-0.06}^{+0.09}$\\ 
     $UVJ$ ($z>3.3$) & 10 & $-0.26_{-0.17}^{+0.15}$ & $0.39_{-0.49}^{+0.43}$ & $0.28_{-0.11}^{+0.13}$\\ 
\hline
$NUVUVJ$ & 12 & $-0.24_{-0.09}^{+0.08}$ & $0.54_{-0.28}^{+0.24}$ & $0.20_{-0.08}^{+0.13}$\\
\hline
vDW14 ($z=2.75$)\tablenotemark{c} & - & $-0.06\pm 0.03$ & $0.79\pm0.07$ & $0.14\pm0.03$
\enddata
\tablenotetext{a}{The number of galaxies used in deriving the best-fit parameters of the size-mass relation.}
\tablenotetext{b}{Their uncertainties are taken from the 16th to 84th percentile of the total chain in the MCMC fitting.}
\tablenotetext{c}{The best-fit parameters of early-type galaxies at $2.5<z<3.0$ reported in \citet{2014ApJ...788...28V}.}
\end{deluxetable}
\subsection{Evolution of the size-mass relation of quiescent galaxies from $z\sim4$}
\par
\begin{figure*}
    \centering
    \includegraphics[width=16cm]{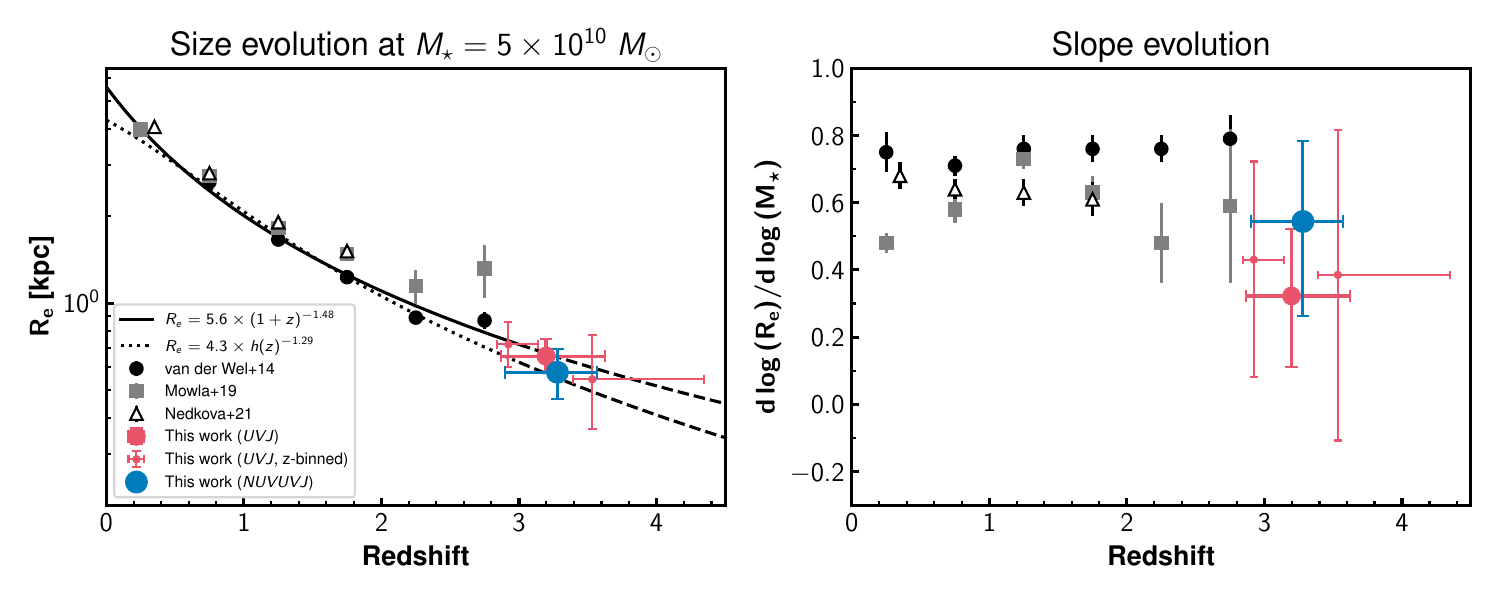}
    \caption{Left panel: Redshift evolution of the effective radius at $M_\star= 5\times10^{10}\, M_\odot$ in the best-fit size-mass relation. Large red and blue circles show the cases for $UVJ$-selected and $NUVUVJ$-selected quiescent galaxies. Small red circles are the case for redshift-sub binned $UVJ$-selected quiescent galaxies. Black circles, gray squares, and open triangles are the lower redshift measurements in \citet{2014ApJ...788...28V}, \citet{2019ApJ...880...57M}, and \citet{2021MNRAS.506..928N}, respectively. Solid and dotted lines are the size evolution function suggested in \citet{2014ApJ...788...28V}, and the dashed lines are their extrapolation at $z>3$. Right panel: Redshift evolution of the slope of the size-mass relation ($\alpha$). The symbols' meanings and the data points' horizontal axis values are the same as in the left panel.}
    \label{fig:param}
\end{figure*}

 Figure \ref{fig:param} shows the evolution of the effective radius at $M_\star= 5\times10^{10}\, M_\odot$ (i.e., $A$) and the slope $\alpha$. In this figure, we compare our values with those of quiescent galaxies at lower redshift ($z<3$) \citep[][]{2014ApJ...788...28V,2019ApJ...880...57M,2021MNRAS.506..928N}. We note that all of the above studies select quiescent galaxies based on the $UVJ$ diagram. We confirm that the effective radius at $M_\star= 5\times10^{10}\, M_\odot$ of this study is smaller than those at $z<3$, implying that the size of the quiescent galaxies monotonically increases since $3<z<4$. Our measured relations are $\sim0.2$ dex below those at $z\sim2$ from the literature and have the smallest amplitude in the studies. Moreover, our measurement is in good agreement with the extrapolation of the redshift evolution of the effective radii suggested in \citet{2014ApJ...788...28V}, which is $R_{e} = 5.6\times(1+z)^{-1.48}\, {\rm kpc}$ or $R_{e} = 4.3\times h(z)^{-1.29}\, {\rm kpc}$. The $UVJ$-selected sample subdivided by redshift is also consistent with these extrapolations.
\par The slope of the size-mass relation of quiescent galaxies does not seem to evolve with redshift. Those drawn from $NUVUVJ$- and $UVJ$-selected quiescent galaxies - also divided in redshift bins - are consistent. Our values are also consistent with those at $z<3$ reported in \citet{2019ApJ...880...57M} and \citet{2021MNRAS.506..928N}. Also, this slope is consistent with the high-mass end slope of the relation in \citet{2022ApJ...925...34C}. However, our values are slightly flatter than those in \citet{2014ApJ...788...28V}, particularly for the $UVJ$-selected sample. The possible origin of this difference will be discussed in Section \ref{sec:dissum}.
\par
\begin{figure}[ht]
    \centering
    \includegraphics[width=8.5cm]{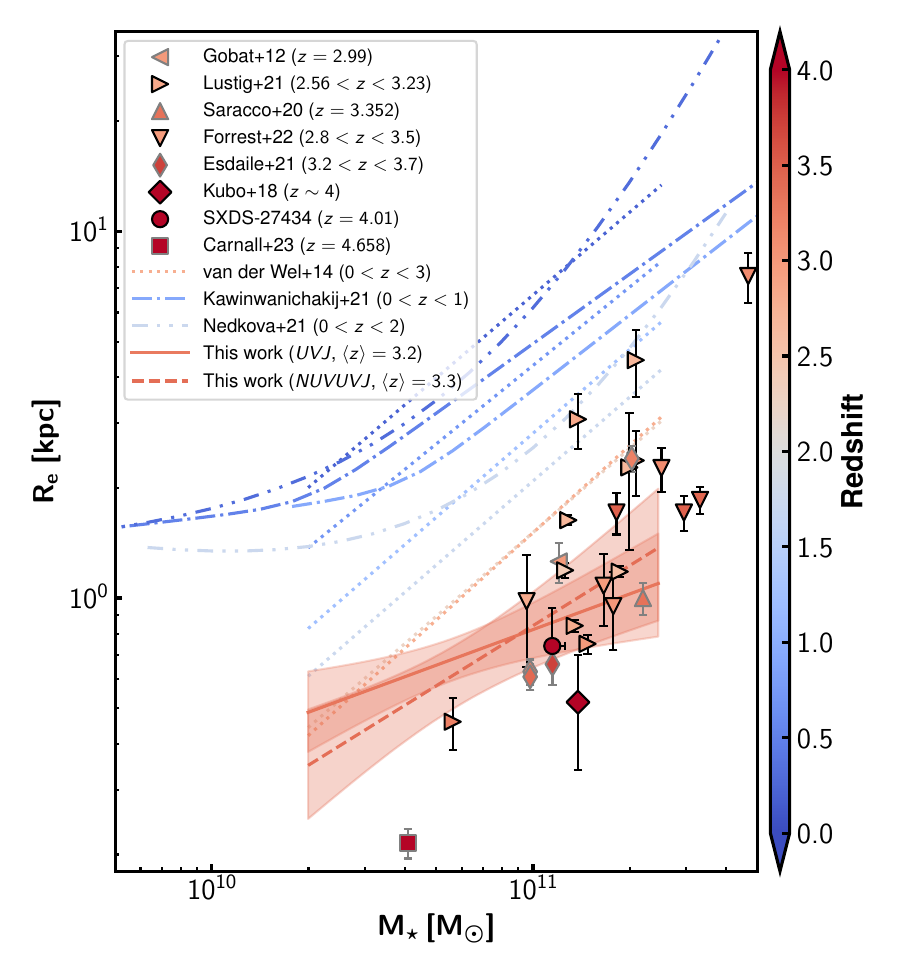}
    \caption{Compilation of the size-mass relation of quiescent galaxies up to $z\sim4$. Those from this study ($z\geq3$) are shown in the solid and dashed lines for $UVJ$ and $NUVUVJ$-selected samples with the uncertainty with the shaded region, respectively. Those at lower redshift are taken from the various literature, \citet{2014ApJ...788...28V} in dotted lines ($0<z<3$), \citet{2021ApJ...921...38K} in dashed-dotted lines ($0<z<1$), and \citet{2021MNRAS.506..928N} in double-dotted line ($0<z<2$), respectively. Their color represents their median redshift. Markers represent the size measurement of individual quiescent galaxies from the literature at $z\geq2.5$ \citep[][]{2012ApJ...759L..44G, 2018ApJ...867....1K, 2020ApJ...905...40S, 2021MNRAS.501.2659L, 2021ApJ...908L..35E,  2022ApJ...938..109F, 2023arXiv230111413C}. They are colored by their redshift, too. We note those from \citet{2022ApJ...938..109F} and \citet{2021MNRAS.501.2659L} are corrected for the wavelength dependence of the size using the empirical relation from \citet{2014ApJ...788...28V}, but the others are not. The wavelength of the used filter is different among them. If the rest-frame wavelength of the size is less than $0.4\, {\rm \mu m}$, the color of their marker edge and error bar is set to be gray. It is set to be black if the rest-frame wavelength of the size is longer than $0.4\, {\rm \mu m}$.}
    \label{fig:sizemass-model}
\end{figure}

 Lastly, Figure \ref{fig:sizemass-model} compares our size-mass relation with the those at lower redshift \citep[][]{2014ApJ...788...28V,2021ApJ...921...38K,2021MNRAS.506..928N} and the size measurements of individual quiescent galaxies at $z\geq2.5$ in the literature \citep[][]{2012ApJ...759L..44G, 2018ApJ...867....1K, 2020ApJ...905...40S,2021MNRAS.501.2659L,  2021ApJ...908L..35E, 2022ApJ...938..109F, 2023arXiv230111413C}. Using the $K_s$-band image of the AO-assisted VLT/HAWK-I, we also obtain the structural parameters of a quiescent galaxy at $z=4.01$ (SXDS-27434) reported in \citet{2019ApJ...885L..34T} and \citet{2020ApJ...889...93V}. Its effective radius is estimated as $r_{\rm e}=0.74^{+0.20}_{-0.09}\, {\rm kpc}$ (see Appendix \ref{app:hawki} for further details). Though the observed wavelength differs among the sample, and many of these values are not corrected for the wavelength dependence of the size \citep[except][]{2021MNRAS.501.2659L, 2022ApJ...938..109F}, they are overall consistent with our measurements. Some of the quiescent galaxies in \citet{2021MNRAS.501.2659L} are larger than our size-mass relation, and this might be due to the redshift difference because of lower redshift for most of them than ours. It should be noted that all $z\geq4$ measurements in \citet{2018ApJ...867....1K} and  \citet{2023arXiv230111413C} are below our size-mass relations, so as the SXDS-27434 at $z=4.01$. This is in agreement with our $z\geq4$ quiescent galaxies (\#2876 at $z=4.146$, \#185 at $z=4.508$, and \#10084 at $z=4.633$) with two-thirds smaller than our relations, which are $0.42\, {\rm kpc}$ at $\log{(M_\star/M_\odot)}=11.0$ on average. Such a systematically small size for the $z\geq4$ sample may imply that the size evolution of quiescent galaxies starts at $z\geq4$.

\section{Morphological Properties} \label{sec:morph}
\subsection{S\'ersic Index and Disk Fraction}\label{sec:n}
\par
\begin{figure}[ht]
    \centering
    \includegraphics[width=8.5cm]{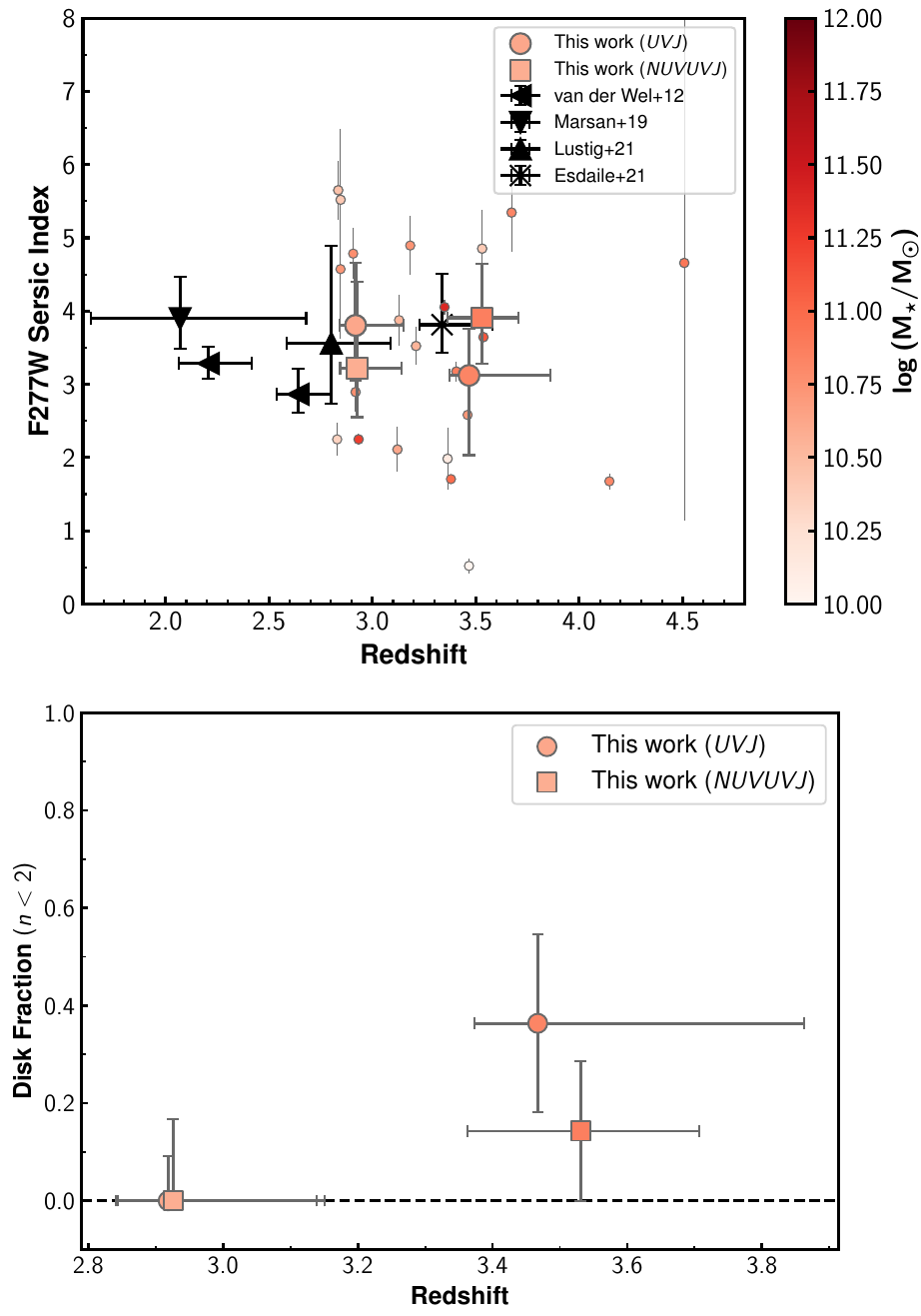}
    \caption{Top panel: S\'ersic index in F277W filter as a function of the redshift. The small colored circles are the individual quiescent galaxies selected by $UVJ$ or $NUVUVJ$. The larger circles and squares are the median of them at $z<3.3$ and $z>3.3$. Black markers show the median of the measurement of quiescent galaxies at $z<3$ in the literature \citep[][]{2012ApJS..203...24V,2019ApJ...871..201M,2021MNRAS.501.2659L,2021ApJ...908L..35E}. The stellar mass range of their sample is $\log{(M_\star/M_\odot)}>10.3$, $\log{(M_\star/M_\odot)}>11.25$, $10.8<\log{(M_\star/M_\odot)}<11.3$, and $11.0<\log{(M_\star/M_\odot)}<11.31$ for \citet{2012ApJS..203...24V}, \citet{2019ApJ...871..201M}, \citet{2021MNRAS.501.2659L}, and \citet{2021ApJ...908L..35E}, respectively. Quiescent galaxies in \citet{2012ApJS..203...24V} are selected in the same manner as \citet{2014ApJ...788...28V}. All of their values are computed by bootstrapping, the same as the value from this work. Bottom panel: Disk fraction, defined as the fraction of sources with F277W S\'ersic index below 2. Circles and squares represent $UVJ$ and $NUVUVJ$-selected quiescent galaxies, respectively.}
    \label{fig:nevo}
\end{figure}

 The S\'ersic index measured in F277W, the filter close to rest-frame $0.5\, {\rm \mu m}$ at $z\sim3-4$, is shown in the top panel of Figure \ref{fig:nevo}. Here, we only focus on the objects with successful measurements of the S\'ersic index (i.e., S\'ersic index is not fixed to $n=4$ in the fitting). The measured S\'ersic indices range from $n=0.5$ to $n=6$, with median values of $n=3.42_{-0.57}^{+0.49}$ and $n=3.61_{-0.47}^{+0.51}$ for $UVJ$-selected and $NUVUVJ$-selected quiescent galaxies, respectively. These values and uncertainties are computed by bootstrapping 5000 times, considering the uncertainty of the S\'ersic index of each galaxy. 
The top panel of Figure \ref{fig:nevo} shows the median of two subsamples divided according to their redshift, as done in Section \ref{sec:sizemass-fit}. They do not show significant redshift evolution in their median values and are in good agreement with lower redshift results \citep[e.g.,][]{2012ApJS..203...24V,2019ApJ...871..201M,2021MNRAS.501.2659L,2021ApJ...908L..35E}. This suggests that most quiescent galaxies have bulge-dominant structures as early as $z\sim4$. 
\par One point that we should mention from our sample is that some quiescent galaxies have a low S\'ersic index as $n\sim0.5-2$. To investigate this further, we next derive the disk fraction, defined as the number of galaxies with $n<2$ divided by the total number of galaxies. These values and their uncertainties are also computed by bootstrapping. The bottom panel of Figure \ref{fig:nevo} shows the disk fraction as a function of redshift. Though both bins at lower redshifts are consistent with zero, the disk fraction increases towards higher redshifts. In particular, the high redshift bin ($z>3.3$) for $UVJ$-selected quiescent galaxies has a non-zero value ($0.36_{-0.18}^{+0.09}$) even considering the $1\sigma$ uncertainty. Our sample suggests that some fraction of quiescent galaxies at high redshift can be disk-dominant morphology, in line with the literature \citep[e.g.,][]{2017Natur.546..510T,2018ApJ...862..126N,2022ApJ...938L..24F}.
\subsection{Axis Ratio}
\par
\begin{figure}[ht]
    \centering
    \includegraphics[width=8.5cm]{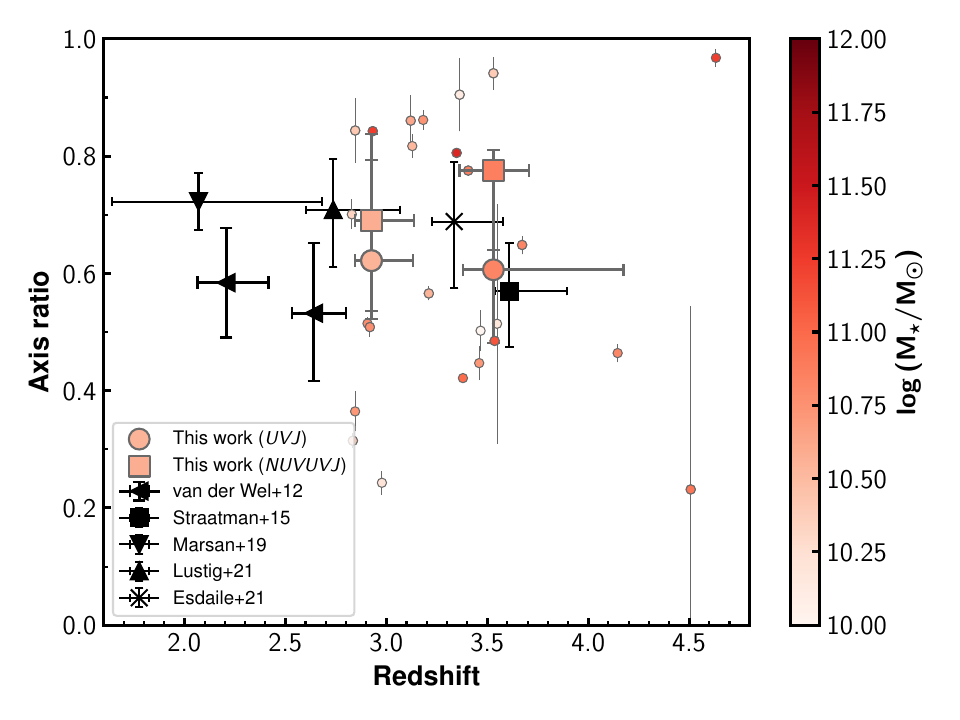}
    \caption{Axis ratio in F277W filter as a function of the redshift. Small colored circles are those of the quiescent galaxies used in this study, and the larger circles and squares are the median of the $UVJ$-selected and $NUVUVJ$-selected samples, respectively. They are subdivided into two samples according to their redshift ($z\leq3.3$ and $z>3.3$). Black markers show the measurement of quiescent galaxies in the literature \citep{2012ApJS..203...24V,2015ApJ...808L..29S,2019ApJ...871..201M,2021MNRAS.501.2659L,2021ApJ...908L..35E}.}
    \label{fig:q}
\end{figure}

 The observed axis ratio of each galaxy projected to the images depends on its inclination angle, but we can trace the intrinsic shape of the sample by taking the median and removing the effect from the inclination angle. We find that the median axis ratio in F277W is $q=0.61_{-0.09}^{+0.12}$ and $q=0.78_{-0.20}^{+0.04}$ for $UVJ$-selected and $NUVUVJ$-selected sample, respectively. These values and uncertainties are again obtained from bootstrapping, as for the median value of the S\'ersic index (Section \ref{sec:n}). The $NUVUVJ$-selected sample has a higher median value than the $UVJ$-selected sample, though with a notably larger uncertainty. Figure \ref{fig:q} shows the individual values of our sample as a function of their redshift and the median value for $2.8< z \leq3.3$ and $3.3<z<4.6$ sample. We do not see any significant dependence of the axis ratio on either the redshift or the stellar mass.
\par Our median values are in good agreement with the measurement of quiescent galaxies at $z<3$ \citep[e.g.,][]{2012ApJS..203...24V,2015ApJ...808L..29S,2021MNRAS.501.2659L} and at $z\sim3-4$ \citep{2015ApJ...808L..29S,2021ApJ...908L..35E}. Coupled with the median S\'ersic index, these values provide further evidence that the majority of quiescent galaxies have bulge-like structures as early as $z\sim4$.

\section{Discussion and Summary} \label{sec:dissum}
\par This study investigates the morphological properties of 26 quiescent galaxies at $3<z<5$ in public $JWST$ fields. Quiescent galaxies are identified by \citet{2023ApJ...947...20V}, selecting targets from all available public $JWST$ data taken during the first three months of operation based on two rest-frame color selections, i.e., the traditional $UVJ$ diagram \citep{2009ApJ...691.1879W} and a novel selection from the rest-frame $NUV$, $U$, $V$, and $J$ band magnitude following \citet{2023arXiv230210934G}. By applying the \textsc{Galfit} software to all NIRCam images of these quiescent galaxies, we derive their effective radius, the S\'ersic index, and the axis ratio at observed $1.15-4.4\, {\rm \mu m}$.
\par Thanks to the measurements in the multiple bands, we derive the rest-frame $0.5\, {\rm \mu m}$ effective radius by considering the wavelength dependence of the effective radius. As those at lower redshift \citep[e.g.,][]{2014ApJ...788...28V,2021ApJ...921...38K,2022ApJ...937L..33S}, quiescent galaxies at $z\geq3$ show a significant negative color gradient in effective radius, i.e., having a more compact size at a longer wavelength. This can be due to the radial dependence of their stellar population. If the inside-out quenching happens, the older stellar population is more concentrated than the younger ones. Since longer wavelength (i.e., $\lambda>0.5\ \mu$m in the rest-frame in our case) sees a more older stellar population, this can cause the observed negative wavelength dependence of the size. Our observed trend implies that correcting for the wavelength dependence of the size measurements is critical for fairly comparing the sizes of quiescent galaxies at different redshifts.
\par We derive the relation of the stellar mass and the effective radius at the rest-frame $0.5\, {\rm \mu m}$ for quiescent galaxies at $3<z<5$ and the best-fit parameters assuming a power-law relation in the stellar mass range of $\log{(M_\star/M_\odot)}>10.3$. These measurements suggest a correlation between the stellar mass and effective radii as early as $z\geq3$ with the power law slope of $0.3-0.5$. In addition, by splitting the $UVJ$-selected sample into two redshift subsamples, we see the possible redshift evolution of its amplitude at $z\geq3$.
\par The effective radii at $M_\star = 5\times 10^{10}\, M_\odot$ from the best-fit are $\sim0.6\, {\rm kpc}$, $\sim0.2$ dex smaller than that of $z\sim2$. This value is the smallest at $0<z<4$. Therefore, we show that the size evolution of quiescent galaxies has started already at $3<z<5$. Our results agree with the redshift evolution in size extrapolated from lower redshift measurements in \citet{2014ApJ...788...28V}.
\par The slope of the size-mass relation does not show the significant redshift evolution in our sample. Though having considerable uncertainty, our measurement shows a flatter slope ($\sim0.3-0.5$) than that of \citet[][$\sim 0.7-0.8$]{2014ApJ...788...28V} in particular for the $UVJ$-selected sample, and close to those of \citet{2019ApJ...880...57M}, \citet{2021MNRAS.506..928N} and the high-mass end slope of the broken power law fit in \citet{2022ApJ...925...34C}. Several explanations exist for the possible difference between our results and \citet{2014ApJ...788...28V}. 
Firstly, this can be due to the increasing fraction of transitional galaxies from star-forming to quiescent, particularly at lower stellar mass. For example, \citet{2021ApJ...915...87S} shows that the median size of green valley galaxies lies between star-forming and quiescent galaxies. If this trend persists at $z\geq3$, such larger galaxies flatten the size-mass relation. Secondly, this can be due to the possible redshift dependence of the stellar mass distribution of the sample, i.e., less low-mass galaxies at higher redshift (see Figure \ref{fig:sizemass-zbin}). Since the stellar mass limit is well below the focused stellar mass range \citep[see the discussion in][]{2023ApJ...947...20V}, this can be due to the small number of the sample or the cosmic variance due to our smaller survey field than that of \citet{2014ApJ...788...28V}. Even though this can not fully explain the difference between our results and \citet{2014ApJ...788...28V} since the slope is still flatter when the sample is split into two redshift subsamples (Figure \ref{fig:sizemass-zbin}), this increase the size at the low stellar mass compared to high mass due to the larger galaxies at lower redshift, leading to the flatter slope. Lastly, the method of collecting the wavelength dependence of the size to derive the rest-frame $0.5\, {\rm \mu m}$ size can make the relation different. We correct it for individual galaxies so as \citet{2021MNRAS.506..928N}, whereas \citet{2014ApJ...788...28V} apply the same correction factor for all galaxies. If the amount of the wavelength dependence of the size is different among galaxies, such as if there is a correlation between the stellar mass and the size gradient like at lower redshift \citet{2021ApJ...921...38K} or if there is a scatter (see Section \ref{sec:cg}), this could make the relation in \citet{2014ApJ...788...28V} different from ours. On the other hand, we have confirmed that only this can not cause this difference either by deriving the size-mass relation using the effective radius at the rest-frame $0.5\, {\rm \mu m}$ from the observed value at F277W and the single median $\gamma$.
\par The high median S\'ersic index ($n\sim 3-4)$ and the axis ratios ($q\sim0.6-0.8$) imply that most quiescent galaxies have bulge-like morphologies. At the same time, we see an increasing trend of the fraction of disk-like morphology toward higher redshift in the $UVJ$-selected sample. The fact that both two spectroscopically confirmed quiescent galaxies at $z>4$ so far have low S\'ersic indexes, i.e., $n={1.23}^{+0.65}_{-0.91}$ for SXDS-27434 (Appendix \ref{app:hawki}) and $n=2.3\pm0.3$ for ones in \citet{2023arXiv230111413C}, also agrees with this trend. These might imply that $3<z<5$ is the epoch when some fraction of quiescent galaxies is in their morphological transformation phase from disk galaxies to elliptical galaxies.
\par This study demonstrates that $JWST$ enables us to explore the structural properties of quiescent galaxies $z>3$. This study serves as a first look at this topic with $JWST$. We only focus on the half-light radii here, but note that future work based on the $JWST$ multi-band imaging will enable us to determine the half-mass radii, as done with $HST$ \citep{2021ApJ...915...87S}. In addition, the wider coverage incoming from new imaging surveys with \textit{JWST}, such as COSMOS-Web \citep{2022arXiv221107865C}, will constrain the structural properties of a significantly larger sample of quiescent galaxies at $z\geq3$. This will reduce the uncertainty of the analytical fitting of the size-mass relation. Also, the systematic comparison with the morphological properties of star-forming galaxies at $z>3$ will be helpful to constrain their quenching mechanism, as done at $z<3$ \citep[e.g.,][]{2017ApJ...839...71F}. Moreover, by increasing the number of quiescent galaxies at $z\geq4$, we can constrain the trend at an even higher redshift and obtain a statistical census into the connection between structural transformation and quenching of the current most distant quiescent galaxies.
\begin{acknowledgments}
\par We appreciate the anonymous referee for helpful comments and suggestions that improved the manuscript. This study is based on the observations associated with programs ERS \#1324, 1345, and 1355; ERO \#2736; GO \#1837 and 2822; GTO \#2738; and COM \#1063. The authors acknowledge the teams and PIs for developing their observing program with a zero-exclusive-access period. Part of the study is based on observations collected at the European Southern Observatory under ESO program 0104.B-0213(A).
This study was supported by JSPS KAKENHI Grant Numbers JP21K03622, JP22J00495, and JP23K13141. OI acknowledges the funding of the French Agence Nationale de la Recherche for the project iMAGE (grant ANR-22-CE31-0007). The Cosmic Dawn Center of Excellence is funded by the Danish National Research Foundation under grant No. 140. GEM acknowledges the Villum Fonden research grant 13160 “Gas to stars, stars to dust: tracing star formation across cosmic time,” grant 37440, “The Hidden Cosmos”. 

\end{acknowledgments}
\facilities{\textit{JWST}}
\software{Astropy \citep{Robitaille2013,Price-Whelan2018}, {\tt emcee} \citep{2013PASP..125..306F}, \textsc{galfit} \citep{2010AJ....139.2097P}, lmfit \citep{newville_matthew_2014_11813}, Matplotlib \citep{Hunter2007}, numba \citep{Lam2015}, numpy \citep{Harris2020}, pandas \citep{Mckinney2010}, photutils \citep{larry_bradley_2022_6825092}, SciPy \citep{2020SciPy-NMeth}, {\tt WebbPSF} \citep{2012SPIE.8442E..3DP,2014SPIE.9143E..3XP}}

\appendix
\section{Selection of stars for smoothing PSFs and an example of smoothed PSF}\label{sec:app1}
\par Stars used to compare with the model are selected based on the following four criteria. (1). Objects with the signal-to-noise ratio of $S/N>200$ in $0.5\arcsec$ aperture magnitude are selected to focus only on those with reliable morphology information. We note that saturated stars are not selected here. (2). Ellipticity, defined as the $(a-b)/(a+b)$, where $a,\ {\rm and}\ b$ are the major and minor axis radius, should be lower than 0.1 to remove the galaxies. (3). Objects on a sequence of unresolved objects are selected based on {\tt FLUX\_RADIUS} from \textsc{sep}. We define the range of the sequence as the $1\sigma$ from the mean of the sequence. (4). If objects have brighter sources within 1\arcsec (2\arcsec) in short (long) wavelength images, they are not used to avoid the impact of the surrounding sources.
\par
\begin{figure*}[ht]
    \centering
    \includegraphics[width=16cm]{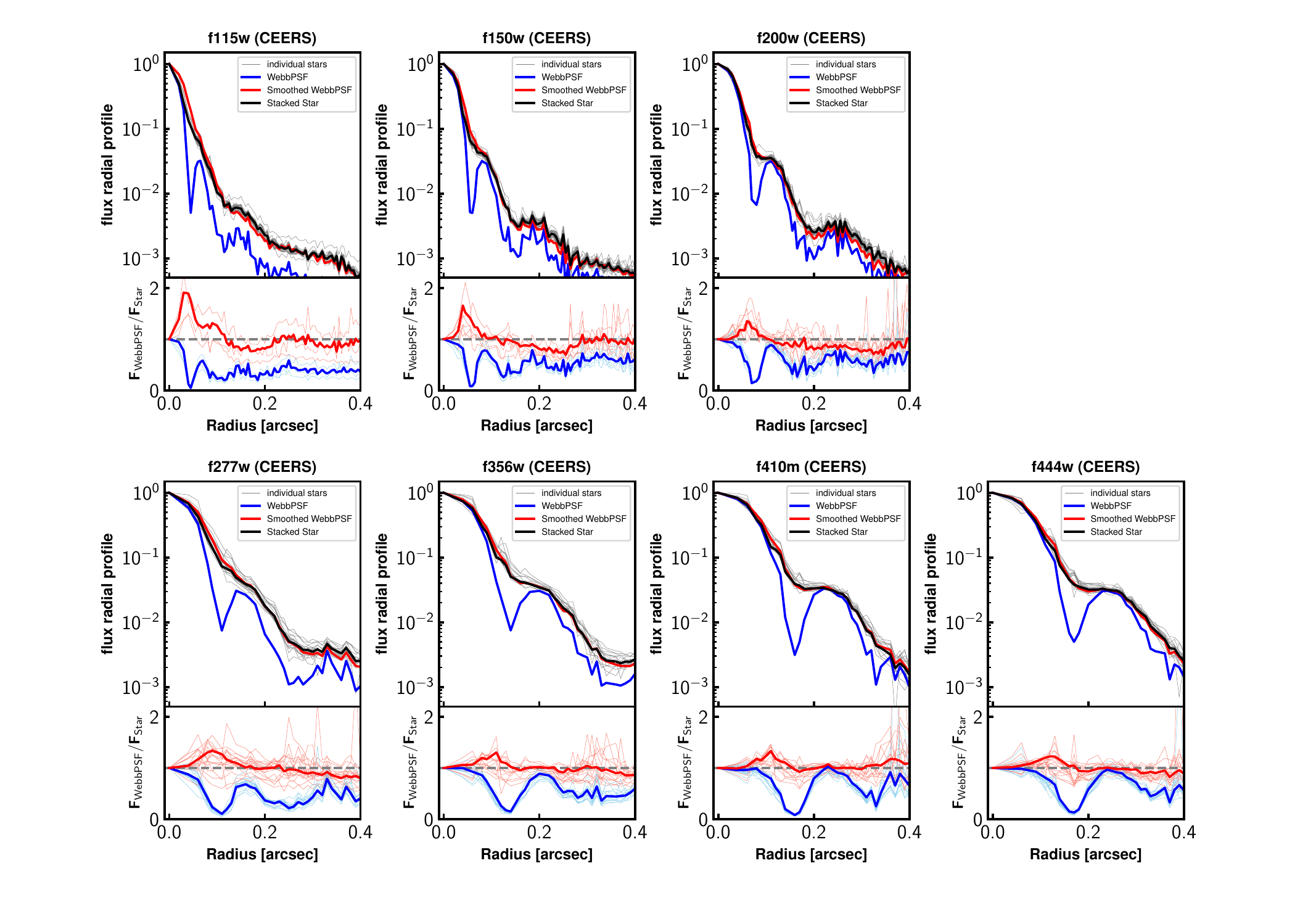}
    \caption{Summary of point spread function of F115W, F150W, F200W, F277W, F356W, F410M and F444W filter in CEERS. Each panel has the radial profile of our smoothed PSF from {\tt WebbPSF} (red), the original PSF from  {\tt WebbPSF} (blue), the median-stacked stars (black), and individual stars (thin gray) at the top. All of them are normalized to have the unity flux at $r=0\arcsec$. The bottom panel shows the ratio of the radial profile between the smoothed PSF (original PSF) and the stacked stars in the red (blue) line. The ratio between those and the individual stars is shown in thinner lines.}
    \label{fig:PSF}
\end{figure*}

 Figure \ref{fig:PSF} shows the radial profile of smoothed PSFs used in this study, the natural stars, and the original PSFs from {\tt WebbPSF} of CEERS as an example. The original PSF from {\tt WebbPSF} is more compact than the natural stars, particularly at shorter wavelengths. The PSFs smoothed by the Moffat function agree mostly well with natural stars in all filters. The median half-width at half maximum of PSFs for each filter among different fields is summarized in Table \ref{tab:hwhm}. The scatter among the values for different fields is 0\%-11\%, particularly smaller values at longer wavelengths. The effective radius at the rest-frame $0.5\, {\rm\mu m}$ of our sample is determined primarily by the size measurement at wavelengths longer than $2\, {\mu m}$ due to the signal-to-noise cut for the magnitude. At that wavelength range, the PSFs are determined homogeneously among fields.
\begin{deluxetable}{cc|cc}
    \tablecaption{Summary of the median half width at half maximum of used PSFs for images of each filter among fields.}\label{tab:hwhm}
    \tablehead{
    \colhead{filter name} & \colhead{wavelength} &\colhead{HWHM} &\colhead{field variation}\\
    \colhead{} & \colhead{(${\rm \mu m}$)} & \colhead{(arcsec)} & \colhead{(\%)}}
    \startdata
        f115w & 1.15 & 0.033 & 6.6 \\
        f150w & 1.5 & 0.035 & 11.1 \\
        f200w & 2.0 & 0.039 & 5.1 \\
        f277w & 2.77 & 0.062 & 4.6 \\
        f356w & 3.56 & 0.069 & 1.5 \\
        f410m & 4.1 & 0.075 & 1.0 \\
        f444w & 4.44 & 0.079 & 0.2 
    \enddata
\end{deluxetable}
\section{Summary of fitting result}\label{app:2}
\par Figure \ref{fig:fit-summary_lowz}, Figure \ref{fig:fit-summary_midz}, and Figure \ref{fig:fit-summary_highz} show the fitting results of all 26 objects which are used in this study. F200W and F277W are selected here because they correspond to the rest-frame $\sim0.5\, {\rm \mu m}$ at redshift of our interest. In addition, Table \ref{tab:2} summarizes the best-fit parameters obtained in fitting F277W images, the effective radius at rest-frame $0.5\, {\rm \mu m}$ and the wavelength dependence of the size derived in Section \ref{sec:cg}. 
\begin{figure*}
    \centering
    \includegraphics[width=17cm]{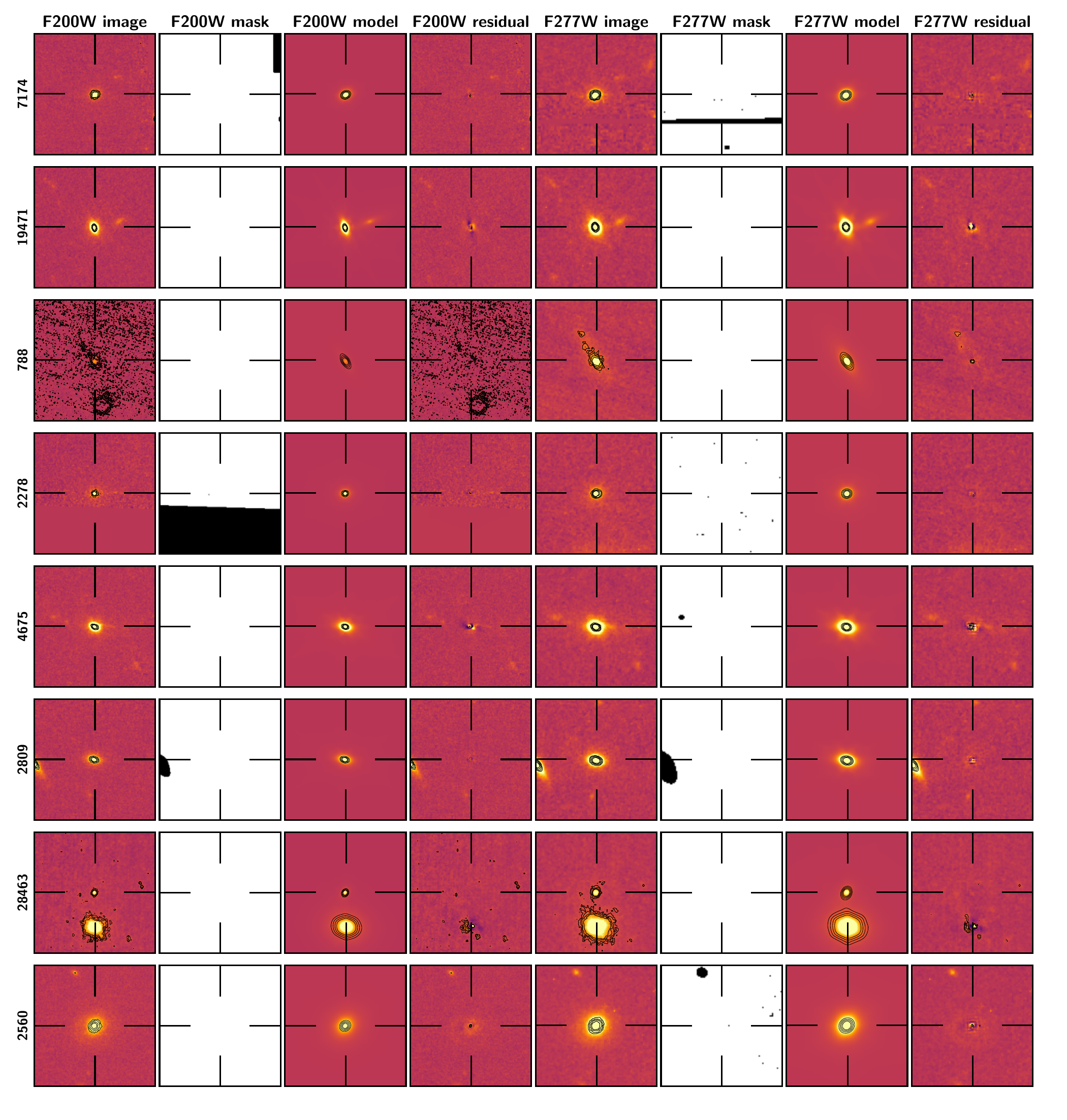}
    \caption{Summary of the fitting result in F200W (left) and F277W (right) images, respectively. For each filter and each object, its image, mask, model, and residual are shown. All are $4.0\arcsec \times 4.0\arcsec$. The images are scaled by their noise ($\pm15\sigma$). The black contours start at $10\%$ of the peak pixel count of each object and increase by $\sqrt{2}$.}
    \label{fig:fit-summary_lowz}
\end{figure*}
\begin{figure*}
    \centering
    \includegraphics[width=17cm]{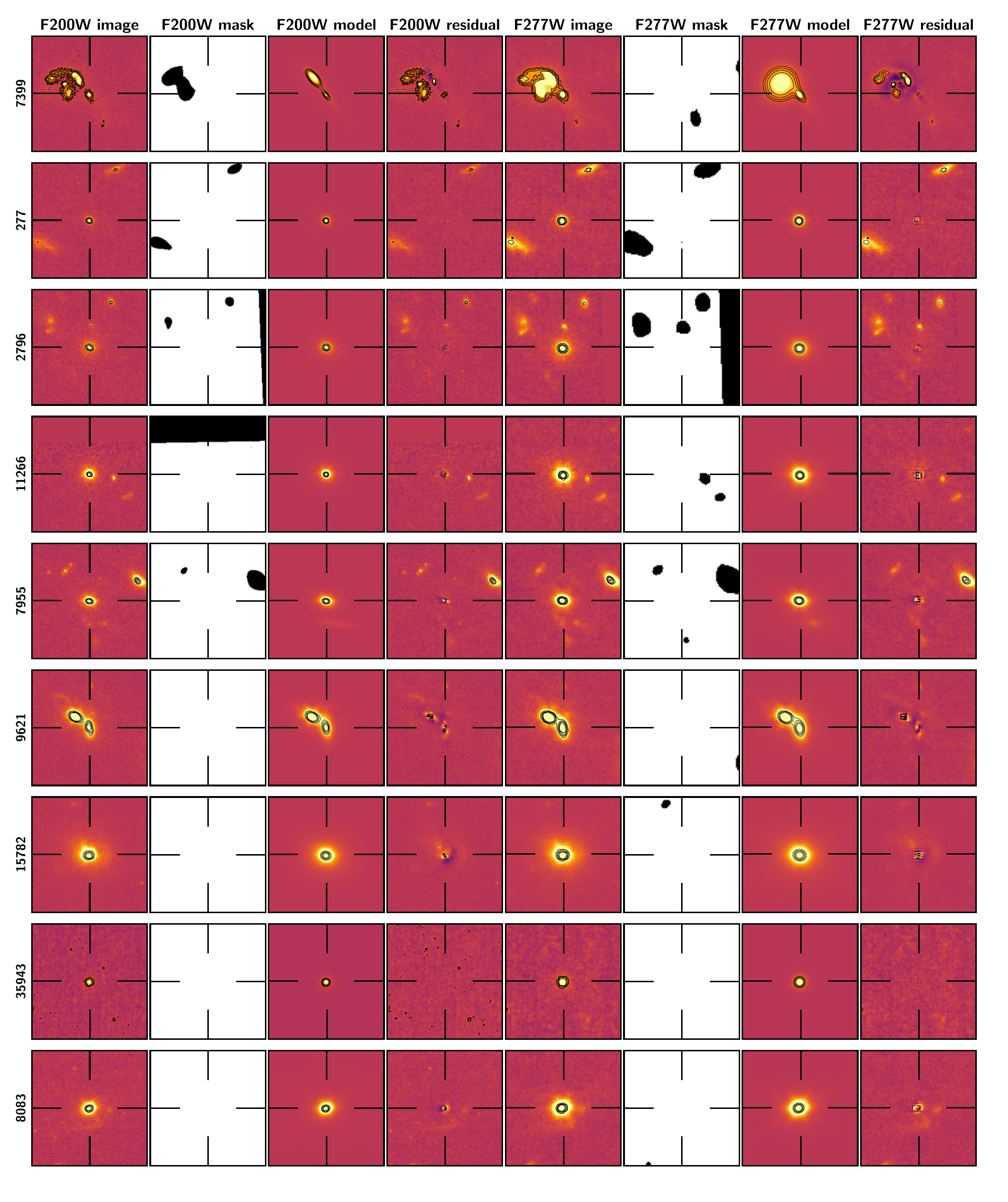}
    \caption{Continued from Figure \ref{fig:fit-summary_lowz}.}
    \label{fig:fit-summary_midz}
\end{figure*}
\begin{figure*}
    \centering
    \includegraphics[width=17cm]{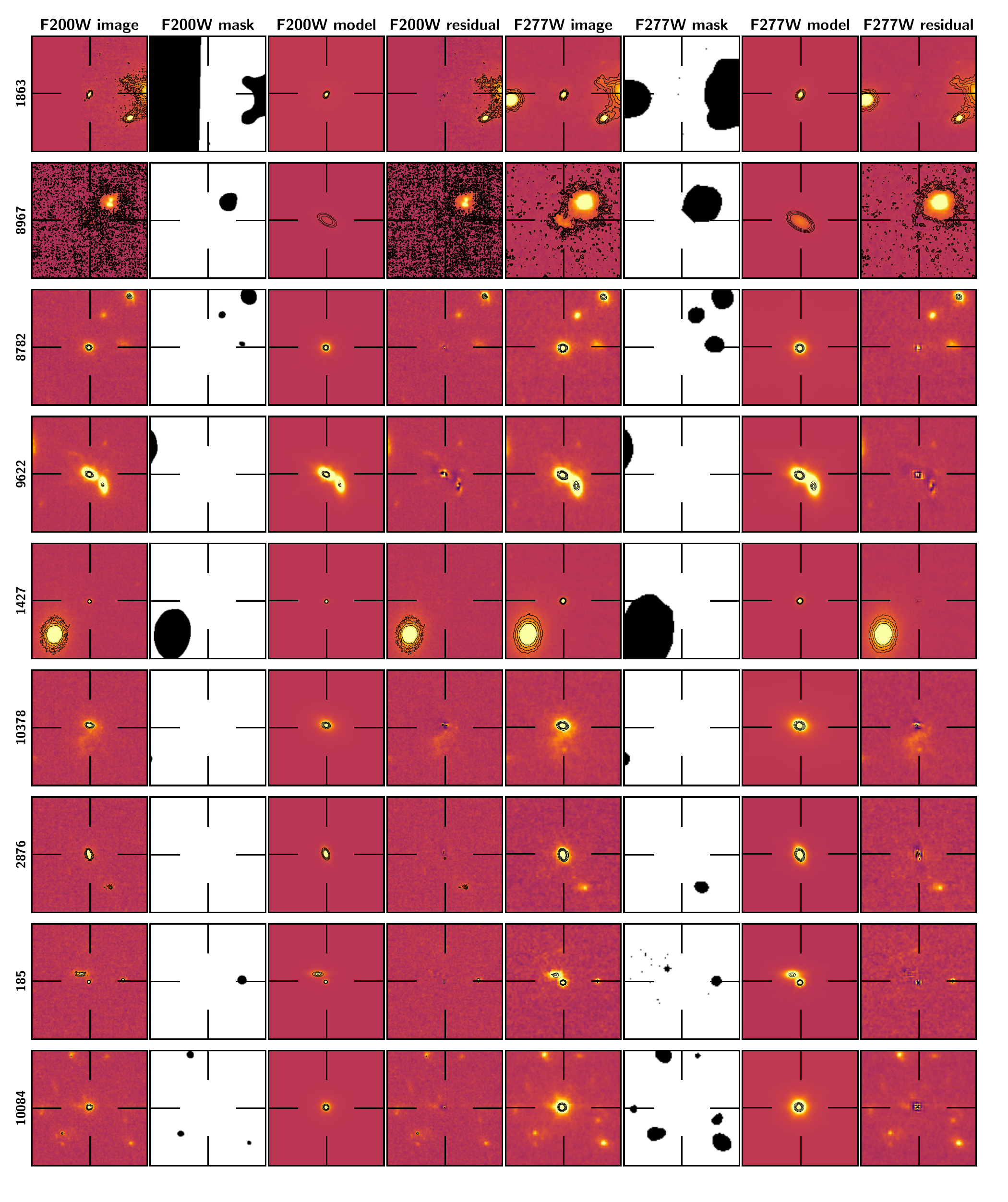}
    \caption{Continued from Figure \ref{fig:fit-summary_lowz}.}
    \label{fig:fit-summary_highz}
\end{figure*}

\movetabledown=7cm
\begin{rotatetable}
\begin{deluxetable}{c|cccccccccccc}
\tablecaption{Summary of \textsc{Galfit} fitting.}\label{tab:2}
\tablehead{
\colhead{ID\tablenotemark{a}} & \colhead{field} &\colhead{$z_{\rm phot}$}\tablenotemark{b} & \colhead{$\log{(M_\star/M_\odot)}$\tablenotemark{b}} & \colhead{$r_{\rm eff, f277w}$} & \colhead{$n_{\rm f277w}$} & \colhead{$q_{\rm f277w}$} & \colhead{$PA_{\rm f277w}$} & \colhead{$m_{\rm f277w}$} & \colhead{$r_{\rm eff, 0.5\mu m}\tablenotemark{c}$} & \colhead{$\gamma$\tablenotemark{d}} & \colhead{$N_{\rm filter}$\tablenotemark{e}} & \colhead{flag\tablenotemark{f}} \\
\colhead{} & \colhead{} & \colhead{} & \colhead{} &\colhead{(kpc, arcsec)}&\colhead{} & \colhead{} & \colhead{(deg)}&\colhead{(mag)} &\colhead{(kpc)}&\colhead{}&\colhead{}&\colhead{}}
\startdata
7174 & CEERS & 2.83 & 10.34 & $0.621\pm0.021$ $(0.079\pm0.003)$ & $2.25\pm0.23$ & $0.701\pm0.025$ & $-61.0\pm3.0$ & $24.25\pm0.02$ & $0.798\pm0.057$ & $-0.57\pm0.11$ & 5 & 0\\
19471 & J1235 & 2.83 & 10.46 & $0.269\pm0.006$ $(0.034\pm0.001\tablenotemark{g})$ & $5.65\pm0.4$ & $0.315\pm0.012$ & $14.0\pm1.0$ & $23.44\pm0.01$ & $0.464\pm0.064$ & $-0.87\pm0.3$ & 5 & 0\\
788 & SGAS1723 & 2.85 & 10.72 & $3.744\pm0.988$ $(0.479\pm0.126)$ & $4.57\pm0.96$ & $0.365\pm0.034$ & $31.0\pm3.0$ & $22.55\pm0.16$ & $4.407\pm3.339$ & $-0.73\pm1.02$ & 3 & 0\\
2278 & PRIMER & 2.85 & 10.41 & $0.722\pm0.094$ $(0.092\pm0.012)$ & $5.52\pm0.97$ & $0.844\pm0.055$ & $-78.0\pm15.0$ & $23.76\pm0.06$ & $0.893\pm0.279$ & $-0.49\pm0.44$ & 4 & 0\\
4675 & CEERS & 2.91 & 10.78 & $0.5\pm0.012$ $(0.064\pm0.002)$ & $4.78\pm0.35$ & $0.514\pm0.01$ & $73.0\pm1.0$ & $22.82\pm0.01$ & $0.602\pm0.032$ & $-0.51\pm0.08$ & 7 & 0\\
2809 & PRIMER & 2.92 & 10.77 & $0.823\pm0.035$ $(0.106\pm0.005)$ & $2.89\pm0.27$ & $0.508\pm0.016$ & $77.0\pm1.0$ & $22.96\pm0.02$ & $0.926\pm0.066$ & $-0.3\pm0.1$ & 6 & 0\\
28463 & QUINTET & 2.93 & 10.4 & $1.023\pm0.2$ $(0.132\pm0.026)$ & $[4.0]$ & $0.626\pm0.08$ & $-32.0\pm8.0$ & $24.49\pm0.12$ & $0.806\pm0.402$ & $0.82\pm0.74$ & 3 & 2\\
2560 & PRIMER & 2.93 & 11.24 & $1.587\pm0.037$ $(0.205\pm0.005)$ & $2.25\pm0.08$ & $0.843\pm0.01$ & $-45.0\pm2.0$ & $21.78\pm0.01$ & $1.836\pm0.113$ & $-0.34\pm0.09$ & 6 & 0\\
7399 & CEERS & 2.98 & 10.21 & $2.508\pm0.202$ $(0.325\pm0.026)$ & $[4.0]$ & $0.243\pm0.02$ & $37.0\pm1.0$ & $24.5\pm0.06$ & $4.882\pm1.185$ & $-1.74\pm0.38$ & 5 & 2\\
277 & PRIMER & 3.12 & 10.62 & $0.336\pm0.012$ $(0.044\pm0.002\tablenotemark{g})$ & $2.11\pm0.31$ & $0.86\pm0.043$ & $35.0\pm11.0$ & $23.39\pm0.01$ & $0.35\pm0.021$ & $-0.07\pm0.11$ & 5 & 0\\
2796 & JWST-NEP-TDF-NRC & 3.13 & 10.53 & $0.809\pm0.047$ $(0.106\pm0.006)$ & $3.88\pm0.35$ & $0.817\pm0.02$ & $75.0\pm3.0$ & $23.38\pm0.03$ & $0.952\pm0.103$ & $-0.65\pm0.17$ & 5 & 0\\
11266 & CEERS & 3.18 & 10.74 & $0.494\pm0.018$ $(0.065\pm0.002)$ & $4.89\pm0.41$ & $0.862\pm0.017$ & $61.0\pm4.0$ & $23.12\pm0.02$ & $0.575\pm0.046$ & $-0.44\pm0.13$ & 6 & 0\\
7955 & CEERS & 3.21 & 10.54 & $0.517\pm0.011$ $(0.069\pm0.001)$ & $3.52\pm0.26$ & $0.566\pm0.012$ & $77.0\pm1.0$ & $23.14\pm0.01$ & $0.591\pm0.046$ & $-0.57\pm0.14$ & 6 & 0\\
9621 & CEERS & 3.38 & 10.98 & $1.272\pm0.015$ $(0.172\pm0.002)$ & $1.7\pm0.05$ & $0.421\pm0.006$ & $9.0\pm0.0$ & $22.9\pm0.01$ & $1.374\pm0.028$ & $-0.37\pm0.04$ & 6 & 0\\
15782 & CEERS & 3.35 & 11.39 & $1.711\pm0.043$ $(0.23\pm0.006)$ & $4.05\pm0.1$ & $0.806\pm0.005$ & $79.0\pm1.0$ & $21.34\pm0.01$ & $1.857\pm0.073$ & $-0.38\pm0.07$ & 7 & 0\\
35943 & QUINTET & 3.36 & 10.12 & $0.708\pm0.067$ $(0.095\pm0.009)$ & $1.98\pm0.43$ & $0.905\pm0.063$ & $-81.0\pm82.0$ & $24.5\pm0.05$ & $0.719\pm0.134$ & $-0.11\pm0.32$ & 3 & 0\\
8083 & CEERS & 3.41 & 10.87 & $0.672\pm0.013$ $(0.091\pm0.002)$ & $3.17\pm0.13$ & $0.775\pm0.009$ & $-71.0\pm1.0$ & $22.49\pm0.01$ & $0.719\pm0.023$ & $-0.31\pm0.06$ & 7 & 0\\
1863 & PRIMER & 3.46 & 10.72 & $0.739\pm0.043$ $(0.101\pm0.006)$ & $2.58\pm0.38$ & $0.447\pm0.029$ & $-28.0\pm2.0$ & $23.81\pm0.04$ & $0.848\pm0.126$ & $-0.41\pm0.26$ & 4 & 1\\
8967 & JWST-NEP-TDF-NRC & 3.47 & 9.77 & $2.214\pm0.153$ $(0.301\pm0.021)$ & $0.52\pm0.1$ & $0.502\pm0.035$ & $60.0\pm3.0$ & $25.46\pm0.09$ & $2.159\pm0.635$ & $-0.37\pm0.56$ & 3 & 0\\
8782 & CEERS & 3.53 & 10.39 & $0.316\pm0.012$ $(0.043\pm0.002\tablenotemark{g})$ & $4.85\pm0.53$ & $0.941\pm0.028$ & $46.0\pm16.0$ & $23.62\pm0.02$ & $0.332\pm0.031$ & $-0.18\pm0.18$ & 6 & 0\\
9622 & CEERS & 3.54 & 11.11 & $0.487\pm0.005$ $(0.067\pm0.001)$ & $3.64\pm0.14$ & $0.485\pm0.006$ & $60.0\pm0.0$ & $22.16\pm0.01$ & $0.532\pm0.031$ & $-0.11\pm0.12$ & 7 & 0\\
1427 & PRIMER & 3.55 & 10.18 & $0.086\pm0.016$ $(0.012\pm0.002)$ & $[4.0]$ & $0.514\pm0.205$ & $-37.0\pm13.0$ & $24.6\pm0.03$ & $0.054\pm0.024$ & $0.67\pm0.95$ & 5 & 3\\
10378 & CEERS & 3.67 & 10.83 & $1.002\pm0.081$ $(0.139\pm0.011)$ & $5.34\pm0.53$ & $0.648\pm0.016$ & $70.0\pm2.0$ & $23.26\pm0.04$ & $1.161\pm0.106$ & $-0.85\pm0.17$ & 6 & 0\\
2876 & CEERS & 4.15 & 10.81 & $0.722\pm0.014$ $(0.105\pm0.002)$ & $1.67\pm0.11$ & $0.464\pm0.016$ & $17.0\pm1.0$ & $23.98\pm0.01$ & $0.743\pm0.019$ & $-0.45\pm0.06$ & 5 & 0\\
185 & PRIMER & 4.51 & 10.91 & $0.033\pm0.035$ $(0.005\pm0.005)$ & $4.66\pm3.52$ & $0.232\pm0.313$ & $-82.0\pm81.0$ & $23.94\pm0.09$ & $0.034\pm0.046$ & $0\pm4\times10^7$ & 3 & 1\\
10084 & CEERS & 4.63 & 11.23 & $0.579\pm0.016$ $(0.089\pm0.002)$ & $[4.0]$ & $0.968\pm0.015$ & $34.0\pm16.0$ & $22.95\pm0.01$ & $0.475\pm0.038$ & $0.15\pm0.21$ & 7 & 2\\
\enddata
\tablenotetext{a}{The IDs are identical to those in \citet{2023ApJ...947...20V}.}
\tablenotetext{b}{They are derived with \textsc{eazy-py} in \citet{2023ApJ...947...20V}.}
\tablenotetext{c}{The effective radius at the rest-frame $0.5\, {\rm \mu m}$ derived in Section \ref{sec:cg}}.
\tablenotetext{d}{The wavelength dependence of the effective radius derived in Section \ref{sec:cg}}.
\tablenotetext{e}{Number of filters used in deriving the $\gamma$ and $r_{\rm eff, 0.5\mu m}$.}
\tablenotemark{f}{The quality of fitting. flag$=0$: good fit. flag$=1$: The \textsc{Galfit} reports an uncertain fit for at least one parameter in one band (i.e., it has parameters marked by *...* in the output file of \textsc{Galfit}). flag$=2$: Its S\'ersic index is fixed as $n=4.0$. flag$=3$: It is flagged as both 1 and 2.}
\tablenotemark{g}{Though the measured effective radii are smaller than the half-width at half maximum of PSFs in F277W, they have larger effective radii than the half-width at half maximum of PSFs in the other filters at shorter wavelength.}
\end{deluxetable}
\end{rotatetable}
\section{Bias and the quality in the fitting for profile with $n=1-3$}
\label{app:mockn13}
\begin{figure}
    \centering
    \includegraphics[width=17cm]{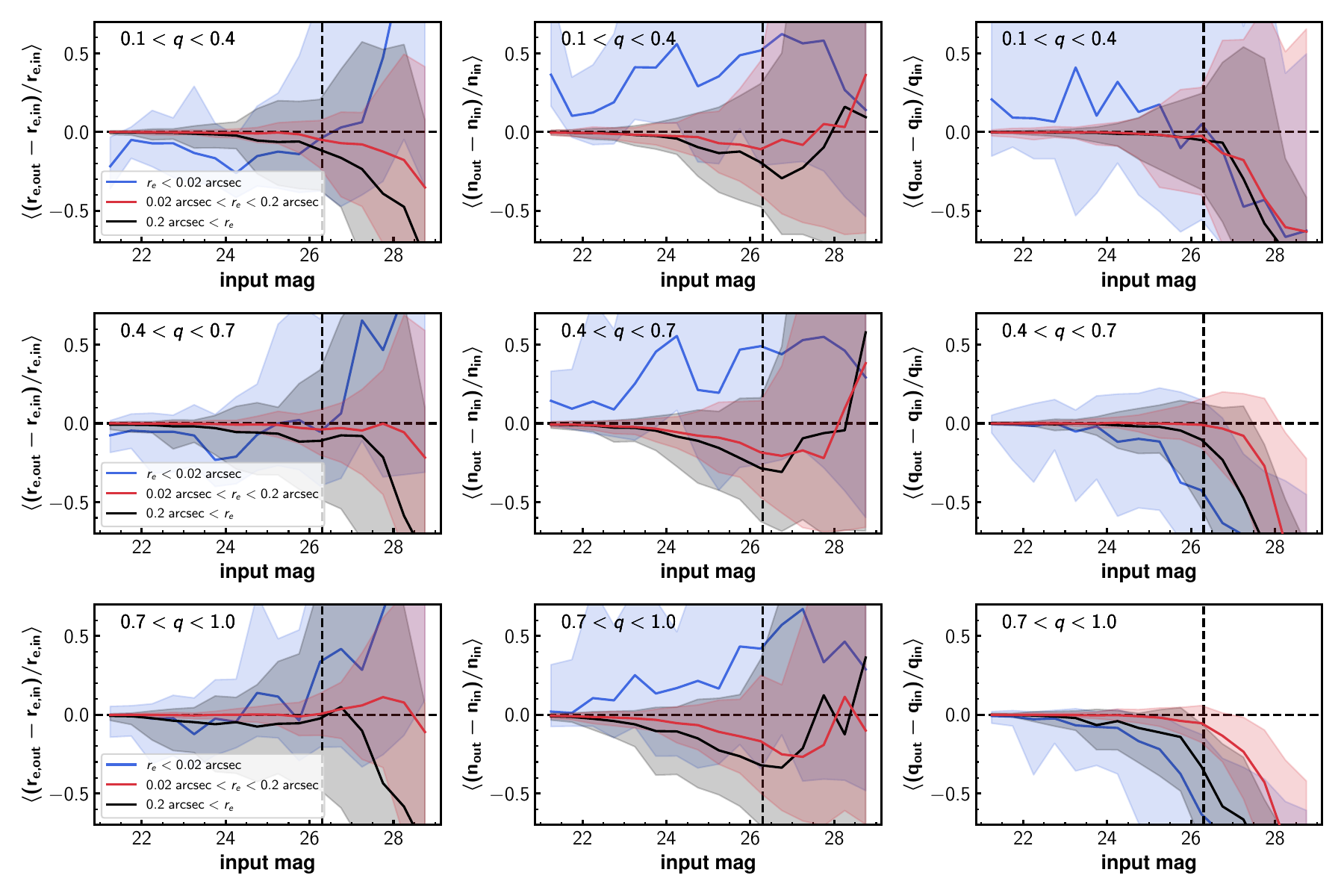}
    \caption{Median relative uncertainties of the estimated parameters (effective radius: left, S\'ersic index: middle, axis ratio: right) derived through mock galaxy images with S\'ersic index of $n=1-3$ as a function of the magnitude. The meanings of each panel and the lines are the same as in Figure \ref{fig:mock}.}
    \label{fig:mock-n13}
\end{figure}

Figure \ref{fig:mock-n13} shows the relative uncertainties of the effective radius, S\'ersic index, and axis ratio for mock S\'ersic profile with $n=1-3$. The trend of the profile with the effective radius of $r_{e}>0.02\arcsec$ is overall consistent with those with $n=3-5$. It is not the case for $r_{e}<0.02\arcsec$, where the effective radius tends to be slightly smaller and the S\'ersic index tends to be larger than the input values.

\section{New size estimate of SXDS-27434 at $z=4.01$}
\label{app:hawki}
\begin{figure*}
    \centering
    \includegraphics[width=\textwidth]{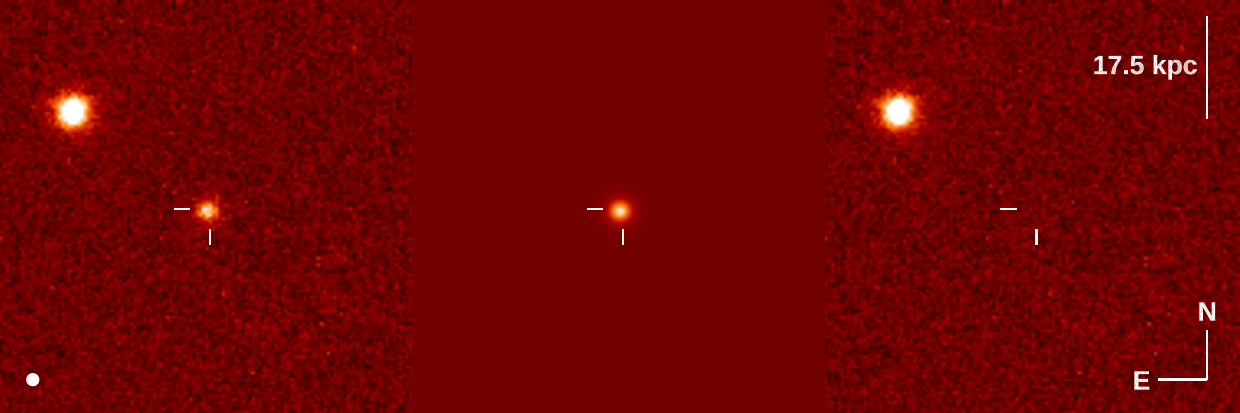}
    \caption{HAWK-I+GRAAL image (\textit{left}), galfit best-fit S\'{e}rsic model (\textit{center}), and residuals (\textit{right}) of SXDS-27434 at $z=4.01$. The size of each cutout is 10\arcsec. The average seeing size is shown by the white circle in the left panel.} 
    \label{fig:hawki} 
\end{figure*}

Here we report the size estimate at rest-frame $\sim4200$ \AA\ of the galaxy SXDS-27434 at $z=4.01$ \citep{2019ApJ...885L..34T, 2020ApJ...889...93V}. This is based on observations with the High Acuity Wide-field K-band Imager \citep[HAWK-I,][]{Pirard2004, Kissler-Patig2008} coupled with the Ground Layer Adaptive Optic system GRAAL \citep{Paufique2010} (Program ID 0104.B-0213, PI: F. Valentino). The target ($\mathrm{R.A., Dec.} = 34.29871, -4.98987$) was observed with the $K_{\rm s}$ filter for $2040\,\mathrm{s}$ on source in service mode on October 22nd, 2019. We applied the standard jitter scheme within a 25\arcsec\ box and obtained 34 offset exposures ($\mathrm{NDIT\times DIT}=6\times10\,\mathrm{s}$). We reduce the data as described in \cite{Brammer2016}\footnote{\url{https://github.com/gbrammer/HAWKI-FF}}. Notably, particular care is given to an optimal sky subtraction with a ``sky flat'' obtained from the median of all the science frames. Finally, the images are ``drizzled'' \citep{Fruchter2002} to a pixel size of $0\farcs08$ and their flux calibration tied to the catalog described in \cite{2019ApJ...885L..34T}. We model the surface brightness of the target as a single S\'{e}rsic profile with galfit. We create a natural PSF by stacking unsaturated stars falling in the HAWK-I footprint, selected in the stellar locus in the \textsc{FLUX\textunderscore RADIUS-MAG\textunderscore AUTO} plane generated with Source Extractor \citep{Bertin1996} and visually inspected. The final average seeing is of $\mathrm{FWHM}=0\farcs34$. The best-fit model and the residuals are shown in Figure \ref{fig:hawki}. We measure an effective semi-major axis of $r_{\rm e}=0.74^{+0.20}_{-0.09}$ kpc and a S\'{e}rsic index $n={1.23}^{+0.65}_{-0.91}$ with an axis ratio $q=0.88^{+0.06}_{-0.26}$. The uncertainties are estimated by bootstrapping $1000\times$ the input cutouts within the rms per pixel measured from the calibrated image and relaunching galfit in the same configuration as for the science measurement. The error bars are computed as the 16-84\% percentiles of the distribution of the parameters in the bootstrapped images. The measurements are robust against the choice of the size of the drizzled pixels ($0\farcs0.8$ or $0\farcs1$) and a fixed or freely varying sky level. The target is resolved, and the radius is well-constrained. This is confirmed by the structured residuals obtained when fitting a single point source in lieu of a S\'{e}rsic profile. The effective radius estimate is consistent with that from $i$-band Hyper-SuprimeCam observations reported in \cite{2019ApJ...885L..34T} ($r_{\rm e}=0.76\pm0.20$ kpc). 

\bibliographystyle{aasjournal}

\begin{thebibliography}{}
\expandafter\ifx\csname natexlab\endcsname\relax\def\natexlab#1{#1}\fi
\providecommand{\url}[1]{\href{#1}{#1}}
\providecommand{\dodoi}[1]{doi:~\href{http://doi.org/#1}{\nolinkurl{#1}}}
\providecommand{\doeprint}[1]{\href{http://ascl.net/#1}{\nolinkurl{http://ascl.net/#1}}}
\providecommand{\doarXiv}[1]{\href{https://arxiv.org/abs/#1}{\nolinkurl{https://arxiv.org/abs/#1}}}

\bibitem[{{Bagley} {et~al.}(2023){Bagley}, {Finkelstein}, {Koekemoer},
  {Ferguson}, {Arrabal Haro}, {Dickinson}, {Kartaltepe}, {Papovich},
  {P{\'e}rez-Gonz{\'a}lez}, {Pirzkal}, {Somerville}, {Willmer}, {Yang}, {Yung},
  {Fontana}, {Grazian}, {Grogin}, {Hirschmann}, {Kewley}, {Kirkpatrick},
  {Kocevski}, {Lotz}, {Medrano}, {Morales}, {Pentericci}, {Ravindranath},
  {Trump}, {Wilkins}, {Calabr{\`o}}, {Cooper}, {Costantin}, {de la Vega},
  {Hilbert}, {Hutchison}, {Larson}, {Lucas}, {McGrath}, {Ryan}, {Wang}, \&
  {Wuyts}}]{2023ApJ...946L..12B}
{Bagley}, M.~B., {Finkelstein}, S.~L., {Koekemoer}, A.~M., {et~al.} 2023,
  \apjl, 946, L12, \dodoi{10.3847/2041-8213/acbb08}

\bibitem[{{Barbary}(2016)}]{2016JOSS....1...58B}
{Barbary}, K. 2016, The Journal of Open Source Software, 1, 58,
  \dodoi{10.21105/joss.00058}

\bibitem[{{Barro} {et~al.}(2013){Barro}, {Faber}, {P{\'e}rez-Gonz{\'a}lez},
  {Koo}, {Williams}, {Kocevski}, {Trump}, {Mozena}, {McGrath}, {van der Wel},
  {Wuyts}, {Bell}, {Croton}, {Ceverino}, {Dekel}, {Ashby}, {Cheung},
  {Ferguson}, {Fontana}, {Fang}, {Giavalisco}, {Grogin}, {Guo}, {Hathi},
  {Hopkins}, {Huang}, {Koekemoer}, {Kartaltepe}, {Lee}, {Newman}, {Porter},
  {Primack}, {Ryan}, {Rosario}, {Somerville}, {Salvato}, \&
  {Hsu}}]{2013ApJ...765..104B}
{Barro}, G., {Faber}, S.~M., {P{\'e}rez-Gonz{\'a}lez}, P.~G., {et~al.} 2013,
  \apj, 765, 104, \dodoi{10.1088/0004-637X/765/2/104}

\bibitem[{{Bertin}(2011)}]{2011ASPC..442..435B}
{Bertin}, E. 2011, in Astronomical Society of the Pacific Conference Series,
  Vol. 442, Astronomical Data Analysis Software and Systems XX, ed. I.~N.
  {Evans}, A.~{Accomazzi}, D.~J. {Mink}, \& A.~H. {Rots}, 435

\bibitem[{{Bertin} \& {Arnouts}(1996)}]{Bertin1996}
{Bertin}, E., \& {Arnouts}, S. 1996, \aaps, 117, 393,
  \dodoi{10.1051/aas:1996164}

\bibitem[{{Bezanson} {et~al.}(2009){Bezanson}, {van Dokkum}, {Tal},
  {Marchesini}, {Kriek}, {Franx}, \& {Coppi}}]{2009ApJ...697.1290B}
{Bezanson}, R., {van Dokkum}, P.~G., {Tal}, T., {et~al.} 2009, \apj, 697, 1290,
  \dodoi{10.1088/0004-637X/697/2/1290}

\bibitem[{Bradley {et~al.}(2022)Bradley, Sip^^c5^^91cz, Robitaille, Tollerud,
  Vin^^c3^^adcius, Deil, Barbary, Wilson, Busko, Donath, G^^c3^^bcnther, Cara,
  Lim, Me^^c3^^9flinger, Conseil, Bostroem, Droettboom, Bray, Bratholm,
  Barentsen, Craig, Rathi, Pascual, Perren, Georgiev, de~Val-Borro, Kerzendorf,
  Bach, Quint, \& Souchereau}]{larry_bradley_2022_6825092}
Bradley, L., Sip^^c5^^91cz, B., Robitaille, T., {et~al.} 2022,
  astropy/photutils: 1.5.0, 1.5.0,  Zenodo, \dodoi{10.5281/zenodo.6825092}

\bibitem[{Brammer(2021)}]{Brammer_eazy-py_2021}
Brammer, G. 2021, {eazy-py}, 0.5.2, \dodoi{10.5281/zenodo.5012704}

\bibitem[{{Brammer} \& {Matharu}(2021)}]{2018zndo...1146904B}
{Brammer}, G., \& {Matharu}, J. 2021, {gbrammer/grizli: Release 2021}, 1.3.2,
  Zenodo,  Zenodo, \dodoi{10.5281/zenodo.1146904}

\bibitem[{{Brammer} {et~al.}(2022){Brammer}, {Strait}, {Matharu}, \&
  {Momcheva}}]{2022zndo...6672538B}
{Brammer}, G., {Strait}, V., {Matharu}, J., \& {Momcheva}, I. 2022, {grizli},
  1.5.0, Zenodo,  Zenodo, \dodoi{10.5281/zenodo.6672538}

\bibitem[{{Brammer} {et~al.}(2008){Brammer}, {van Dokkum}, \&
  {Coppi}}]{2008ApJ...686.1503B}
{Brammer}, G.~B., {van Dokkum}, P.~G., \& {Coppi}, P. 2008, \apj, 686, 1503,
  \dodoi{10.1086/591786}

\bibitem[{{Brammer} {et~al.}(2016){Brammer}, {Marchesini}, {Labb{\'e}},
  {Spitler}, {Lange-Vagle}, {Barker}, {Tanaka}, {Fontana}, {Galametz},
  {Ferr{\'e}-Mateu}, {Kodama}, {Lundgren}, {Martis}, {Muzzin}, {Stefanon},
  {Toft}, {van der Wel}, {Vulcani}, \& {Whitaker}}]{Brammer2016}
{Brammer}, G.~B., {Marchesini}, D., {Labb{\'e}}, I., {et~al.} 2016, \apjs, 226,
  6, \dodoi{10.3847/0067-0049/226/1/6}

\bibitem[{{Carnall} {et~al.}(2023{\natexlab{a}}){Carnall}, {McLeod}, {McLure},
  {Dunlop}, {Begley}, {Cullen}, {Donnan}, {Hamadouche}, {Jewell}, {Jones},
  {Pollock}, \& {Wild}}]{2023MNRAS.520.3974C}
{Carnall}, A.~C., {McLeod}, D.~J., {McLure}, R.~J., {et~al.}
  2023{\natexlab{a}}, \mnras, 520, 3974, \dodoi{10.1093/mnras/stad369}

\bibitem[{{Carnall} {et~al.}(2023{\natexlab{b}}){Carnall}, {McLure}, {Dunlop},
  {McLeod}, {Wild}, {Cullen}, {Magee}, {Begley}, {Cimatti}, {Donnan},
  {Hamadouche}, {Jewell}, \& {Walker}}]{2023arXiv230111413C}
{Carnall}, A.~C., {McLure}, R.~J., {Dunlop}, J.~S., {et~al.}
  2023{\natexlab{b}}, arXiv e-prints, arXiv:2301.11413,
  \dodoi{10.48550/arXiv.2301.11413}

\bibitem[{{Carollo} {et~al.}(2013){Carollo}, {Bschorr}, {Renzini}, {Lilly},
  {Capak}, {Cibinel}, {Ilbert}, {Onodera}, {Scoville}, {Cameron}, {Mobasher},
  {Sanders}, \& {Taniguchi}}]{2013ApJ...773..112C}
{Carollo}, C.~M., {Bschorr}, T.~J., {Renzini}, A., {et~al.} 2013, \apj, 773,
  112, \dodoi{10.1088/0004-637X/773/2/112}

\bibitem[{{Casey} {et~al.}(2022){Casey}, {Kartaltepe}, {Drakos}, {Franco},
  {Harish}, {Paquereau}, {Ilbert}, {Rose}, {Cox}, {Nightingale}, {Robertson},
  {Silverman}, {Koekemoer}, {Massey}, {McCracken}, {Rhodes}, {Akins},
  {Amvrosiadis}, {Arango-Toro}, {Bagley}, {Bongiorno}, {Capak}, {Champagne},
  {Chartab}, {Chavez Ortiz}, {Chworowsky}, {Cooke}, {Cooper}, {Darvish},
  {Ding}, {Faisst}, {Finkelstein}, {Fujimoto}, {Gentile}, {Gillman}, {Gould},
  {Gozaliasl}, {Hayward}, {He}, {Hemmati}, {Hirschmann}, {Jahnke}, {Jin},
  {Khostovan}, {Kokorev}, {Lambrides}, {Laigle}, {Larson}, {Leung}, {Liu},
  {Liaudat}, {Long}, {Magdis}, {Mahler}, {Mainieri}, {Manning}, {Maraston},
  {Martin}, {McCleary}, {McKinney}, {McPartland}, {Mobasher}, {Pattnaik},
  {Renzini}, {Rich}, {Sanders}, {Sattari}, {Scognamiglio}, {Scoville}, {Sheth},
  {Shuntov}, {Sparre}, {Suzuki}, {Talia}, {Toft}, {Trakhtenbrot}, {Urry},
  {Valentino}, {Vanderhoof}, {Vardoulaki}, {Weaver}, {Whitaker}, {Wilkins},
  {Yang}, \& {Zavala}}]{2022arXiv221107865C}
{Casey}, C.~M., {Kartaltepe}, J.~S., {Drakos}, N.~E., {et~al.} 2022, arXiv
  e-prints, arXiv:2211.07865, \dodoi{10.48550/arXiv.2211.07865}

\bibitem[{{Cassata} {et~al.}(2013){Cassata}, {Giavalisco}, {Williams}, {Guo},
  {Lee}, {Renzini}, {Ferguson}, {Faber}, {Barro}, {McIntosh}, {Lu}, {Bell},
  {Koo}, {Papovich}, {Ryan}, {Conselice}, {Grogin}, {Koekemoer}, \&
  {Hathi}}]{2013ApJ...775..106C}
{Cassata}, P., {Giavalisco}, M., {Williams}, C.~C., {et~al.} 2013, \apj, 775,
  106, \dodoi{10.1088/0004-637X/775/2/106}

\bibitem[{{Chabrier}(2003)}]{2003PASP..115..763C}
{Chabrier}, G. 2003, \pasp, 115, 763, \dodoi{10.1086/376392}

\bibitem[{{Cutler} {et~al.}(2022){Cutler}, {Whitaker}, {Mowla}, {Brammer}, {van
  der Wel}, {Marchesini}, {van Dokkum}, {Momcheva}, {Song}, {Akhshik},
  {Nelson}, {Bezanson}, {Franx}, {Kriek}, {Lange-Vagle}, {Leja}, {MacKenty},
  {Muzzin}, \& {Shipley}}]{2022ApJ...925...34C}
{Cutler}, S.~E., {Whitaker}, K.~E., {Mowla}, L.~A., {et~al.} 2022, \apj, 925,
  34, \dodoi{10.3847/1538-4357/ac341c}

\bibitem[{{D'Eugenio} {et~al.}(2021){D'Eugenio}, {Daddi}, {Gobat},
  {Strazzullo}, {Lustig}, {Delvecchio}, {Jin}, {Cimatti}, \&
  {Onodera}}]{2021A&A...653A..32D}
{D'Eugenio}, C., {Daddi}, E., {Gobat}, R., {et~al.} 2021, \aap, 653, A32,
  \dodoi{10.1051/0004-6361/202040067}

\bibitem[{{Ding} {et~al.}(2022){Ding}, {Silverman}, \&
  {Onoue}}]{2022ApJ...939L..28D}
{Ding}, X., {Silverman}, J.~D., \& {Onoue}, M. 2022, \apjl, 939, L28,
  \dodoi{10.3847/2041-8213/ac9c02}

\bibitem[{{Esdaile} {et~al.}(2021){Esdaile}, {Glazebrook}, {Labb{\'e}},
  {Taylor}, {Schreiber}, {Nanayakkara}, {Kacprzak}, {Oesch}, {Tran},
  {Papovich}, {Spitler}, \& {Straatman}}]{2021ApJ...908L..35E}
{Esdaile}, J., {Glazebrook}, K., {Labb{\'e}}, I., {et~al.} 2021, \apjl, 908,
  L35, \dodoi{10.3847/2041-8213/abe11e}

\bibitem[{{Faisst} {et~al.}(2017){Faisst}, {Carollo}, {Capak}, {Tacchella},
  {Renzini}, {Ilbert}, {McCracken}, \& {Scoville}}]{2017ApJ...839...71F}
{Faisst}, A.~L., {Carollo}, C.~M., {Capak}, P.~L., {et~al.} 2017, \apj, 839,
  71, \dodoi{10.3847/1538-4357/aa697a}

\bibitem[{{Foreman-Mackey} {et~al.}(2013){Foreman-Mackey}, {Hogg}, {Lang}, \&
  {Goodman}}]{2013PASP..125..306F}
{Foreman-Mackey}, D., {Hogg}, D.~W., {Lang}, D., \& {Goodman}, J. 2013, \pasp,
  125, 306, \dodoi{10.1086/670067}

\bibitem[{{Forrest} {et~al.}(2020){Forrest}, {Marsan}, {Annunziatella},
  {Wilson}, {Muzzin}, {Marchesini}, {Cooper}, {Chan}, {McConachie}, {Gomez},
  {Kado-Fong}, {La Barbera}, {Lange-Vagle}, {Nantais}, {Nonino}, {Saracco},
  {Stefanon}, \& {van der Burg}}]{2020ApJ...903...47F}
{Forrest}, B., {Marsan}, Z.~C., {Annunziatella}, M., {et~al.} 2020, \apj, 903,
  47, \dodoi{10.3847/1538-4357/abb819}

\bibitem[{{Forrest} {et~al.}(2022){Forrest}, {Wilson}, {Muzzin}, {Marchesini},
  {Cooper}, {Marsan}, {Annunziatella}, {McConachie}, {Zaidi}, {Gomez}, {Urbano
  Stawinski}, {Chang}, {de Lucia}, {La Barbera}, {Lubin}, {Nantais},
  {Pe{\~n}a}, {Saracco}, {Surace}, \& {Stefanon}}]{2022ApJ...938..109F}
{Forrest}, B., {Wilson}, G., {Muzzin}, A., {et~al.} 2022, \apj, 938, 109,
  \dodoi{10.3847/1538-4357/ac8747}

\bibitem[{{Fruchter} \& {Hook}(2002)}]{Fruchter2002}
{Fruchter}, A.~S., \& {Hook}, R.~N. 2002, \pasp, 114, 144,
  \dodoi{10.1086/338393}

\bibitem[{{Fudamoto} {et~al.}(2022){Fudamoto}, {Inoue}, \&
  {Sugahara}}]{2022ApJ...938L..24F}
{Fudamoto}, Y., {Inoue}, A.~K., \& {Sugahara}, Y. 2022, \apjl, 938, L24,
  \dodoi{10.3847/2041-8213/ac982b}

\bibitem[{{Gillman} {et~al.}(2023){Gillman}, {Gullberg}, {Brammer}, {Vijayan},
  {Lee}, {Bl{\'a}nquez}, {Brinch}, {Greve}, {Jermann}, {Jin}, {Kokorev}, {Liu},
  {Magdis}, {Rizzo}, \& {Valentino}}]{2023arXiv230317246G}
{Gillman}, S., {Gullberg}, B., {Brammer}, G., {et~al.} 2023, arXiv e-prints,
  arXiv:2303.17246, \dodoi{10.48550/arXiv.2303.17246}

\bibitem[{{Glazebrook} {et~al.}(2017){Glazebrook}, {Schreiber}, {Labb{\'e}},
  {Nanayakkara}, {Kacprzak}, {Oesch}, {Papovich}, {Spitler}, {Straatman},
  {Tran}, \& {Yuan}}]{2017Natur.544...71G}
{Glazebrook}, K., {Schreiber}, C., {Labb{\'e}}, I., {et~al.} 2017, \nat, 544,
  71, \dodoi{10.1038/nature21680}

\bibitem[{{Gobat} {et~al.}(2012){Gobat}, {Strazzullo}, {Daddi}, {Onodera},
  {Renzini}, {B{\'e}thermin}, {Dickinson}, {Carollo}, \&
  {Cimatti}}]{2012ApJ...759L..44G}
{Gobat}, R., {Strazzullo}, V., {Daddi}, E., {et~al.} 2012, \apjl, 759, L44,
  \dodoi{10.1088/2041-8205/759/2/L44}

\bibitem[{{G{\'o}mez-Guijarro} {et~al.}(2023){G{\'o}mez-Guijarro}, {Magnelli},
  {Elbaz}, {Wuyts}, {Daddi}, {Le Bail}, {Giavalisco}, {Dickinson},
  {P{\'e}rez-Gonz{\'a}lez}, {Arrabal Haro}, {Bagley}, {Bisigello}, {Buat},
  {Burgarella}, {Calabr{\`o}}, {Casey}, {Cheng}, {Ciesla}, {Dekel}, {Ferguson},
  {Finkelstein}, {Franco}, {Grogin}, {Holwerda}, {Jin}, {Kartaltepe},
  {Koekemoer}, {Kokorev}, {Long}, {Lucas}, {Magdis}, {Papovich}, {Pirzkal},
  {Seill{\'e}}, {Tacchella}, {Tarrasse}, {Valentino}, {de la Vega}, {Wilkins},
  {Xiao}, \& {Yung}}]{2023arXiv230408517G}
{G{\'o}mez-Guijarro}, C., {Magnelli}, B., {Elbaz}, D., {et~al.} 2023, arXiv
  e-prints, arXiv:2304.08517, \dodoi{10.48550/arXiv.2304.08517}

\bibitem[{Gould(2023)}]{Gould_GMM-quiescent_2023}
Gould, K. M.~L. 2023, {GMM-quiescent}, 1.0.0, \dodoi{10.5281/zenodo.7712937}

\bibitem[{{Gould} {et~al.}(2023){Gould}, {Brammer}, {Valentino}, {Whitaker},
  {Weaver}, {Lagos}, {Rizzo}, {Franco}, {Hseih}, {Ilbert}, {Jin}, {Magdis},
  {McCracken}, {Mobasher}, {Shuntov}, {Steinhardt}, {Strait}, \&
  {Toft}}]{2023arXiv230210934G}
{Gould}, K. M.~L., {Brammer}, G., {Valentino}, F., {et~al.} 2023, arXiv
  e-prints, arXiv:2302.10934, \dodoi{10.48550/arXiv.2302.10934}

\bibitem[{Harris {et~al.}(2020)Harris, Millman, van~der Walt, Gommers,
  Virtanen, Cournapeau, Wieser, Taylor, Berg, Smith, Kern, Picus, Hoyer, van
  Kerkwijk, Brett, Haldane, del R{\'{i}}o, Wiebe, Peterson,
  G{\'{e}}rard-Marchant, Sheppard, Reddy, Weckesser, Abbasi, Gohlke, \&
  Oliphant}]{Harris2020}
Harris, C.~R., Millman, K.~J., van~der Walt, S.~J., {et~al.} 2020, Nature, 585,
  357, \dodoi{10.1038/s41586-020-2649-2}

\bibitem[{Hunter(2007)}]{Hunter2007}
Hunter, J.~D. 2007, Computing in Science and Engineering, 9, 90,
  \dodoi{10.1109/MCSE.2007.55}

\bibitem[{{Kawinwanichakij} {et~al.}(2021){Kawinwanichakij}, {Silverman},
  {Ding}, {George}, {Damjanov}, {Sawicki}, {Tanaka}, {Taranu}, {Birrer},
  {Huang}, {Li}, {Onodera}, {Shibuya}, \& {Yasuda}}]{2021ApJ...921...38K}
{Kawinwanichakij}, L., {Silverman}, J.~D., {Ding}, X., {et~al.} 2021, \apj,
  921, 38, \dodoi{10.3847/1538-4357/ac1f21}

\bibitem[{{Kissler-Patig} {et~al.}(2008){Kissler-Patig}, {Pirard}, {Casali},
  {Moorwood}, {Ageorges}, {Alves de Oliveira}, {Baksai}, {Bedin}, {Bendek},
  {Biereichel}, {Delabre}, {Dorn}, {Esteves}, {Finger}, {Gojak}, {Huster},
  {Jung}, {Kiekebush}, {Klein}, {Koch}, {Lizon}, {Mehrgan}, {Petr-Gotzens},
  {Pritchard}, {Selman}, \& {Stegmeier}}]{Kissler-Patig2008}
{Kissler-Patig}, M., {Pirard}, J.~F., {Casali}, M., {et~al.} 2008, \aap, 491,
  941, \dodoi{10.1051/0004-6361:200809910}

\bibitem[{{Kocevski} {et~al.}(2018){Kocevski}, {Hasinger}, {Brightman},
  {Nandra}, {Georgakakis}, {Cappelluti}, {Civano}, {Li}, {Li}, {Aird},
  {Alexander}, {Almaini}, {Brusa}, {Buchner}, {Comastri}, {Conselice},
  {Dickinson}, {Finoguenov}, {Gilli}, {Koekemoer}, {Miyaji}, {Mullaney},
  {Papovich}, {Rosario}, {Salvato}, {Silverman}, {Somerville}, \&
  {Ueda}}]{2018ApJS..236...48K}
{Kocevski}, D.~D., {Hasinger}, G., {Brightman}, M., {et~al.} 2018, \apjs, 236,
  48, \dodoi{10.3847/1538-4365/aab9b4}

\bibitem[{{Kubo} {et~al.}(2018){Kubo}, {Tanaka}, {Yabe}, {Toft}, {Stockmann},
  \& {G{\'o}mez-Guijarro}}]{2018ApJ...867....1K}
{Kubo}, M., {Tanaka}, M., {Yabe}, K., {et~al.} 2018, \apj, 867, 1,
  \dodoi{10.3847/1538-4357/aae3e8}

\bibitem[{Lam {et~al.}(2015)Lam, Pitrou, \& Seibert}]{Lam2015}
Lam, S.~K., Pitrou, A., \& Seibert, S. 2015, in Proceedings of the Second
  Workshop on the LLVM Compiler Infrastructure in HPC - LLVM '15 (New York, New
  York, USA: ACM Press), 1--6, \dodoi{10.1145/2833157.2833162}

\bibitem[{{Lustig} {et~al.}(2021){Lustig}, {Strazzullo}, {D'Eugenio}, {Daddi},
  {Pannella}, {Renzini}, {Cimatti}, {Gobat}, {Jin}, {Mohr}, \&
  {Onodera}}]{2021MNRAS.501.2659L}
{Lustig}, P., {Strazzullo}, V., {D'Eugenio}, C., {et~al.} 2021, \mnras, 501,
  2659, \dodoi{10.1093/mnras/staa3766}

\bibitem[{{Marchesini} {et~al.}(2009){Marchesini}, {van Dokkum}, {F{\"o}rster
  Schreiber}, {Franx}, {Labb{\'e}}, \& {Wuyts}}]{2009ApJ...701.1765M}
{Marchesini}, D., {van Dokkum}, P.~G., {F{\"o}rster Schreiber}, N.~M., {et~al.}
  2009, \apj, 701, 1765, \dodoi{10.1088/0004-637X/701/2/1765}

\bibitem[{{Marsan} {et~al.}(2019){Marsan}, {Marchesini}, {Muzzin}, {Brammer},
  {Bezanson}, {Franx}, {Labb{\'e}}, {Lundgren}, {Rudnick}, {Stefanon}, {van
  Dokkum}, {Wake}, \& {Whitaker}}]{2019ApJ...871..201M}
{Marsan}, Z.~C., {Marchesini}, D., {Muzzin}, A., {et~al.} 2019, \apj, 871, 201,
  \dodoi{10.3847/1538-4357/aaf808}

\bibitem[{McKinney(2010)}]{Mckinney2010}
McKinney, W. 2010, {Data Structures for Statistical Computing in Python}, Tech.
  rep., \dodoi{10.25080/majora-92bf1922-00a}

\bibitem[{{Mosleh} {et~al.}(2017){Mosleh}, {Tacchella}, {Renzini}, {Carollo},
  {Molaeinezhad}, {Onodera}, {Khosroshahi}, \& {Lilly}}]{2017ApJ...837....2M}
{Mosleh}, M., {Tacchella}, S., {Renzini}, A., {et~al.} 2017, \apj, 837, 2,
  \dodoi{10.3847/1538-4357/aa5f14}

\bibitem[{{Moster} {et~al.}(2011){Moster}, {Somerville}, {Newman}, \&
  {Rix}}]{2011ApJ...731..113M}
{Moster}, B.~P., {Somerville}, R.~S., {Newman}, J.~A., \& {Rix}, H.-W. 2011,
  \apj, 731, 113, \dodoi{10.1088/0004-637X/731/2/113}

\bibitem[{{Mowla} {et~al.}(2019{\natexlab{a}}){Mowla}, {van der Wel}, {van
  Dokkum}, \& {Miller}}]{2019ApJ...872L..13M}
{Mowla}, L., {van der Wel}, A., {van Dokkum}, P., \& {Miller}, T.~B.
  2019{\natexlab{a}}, \apjl, 872, L13, \dodoi{10.3847/2041-8213/ab0379}

\bibitem[{{Mowla} {et~al.}(2019{\natexlab{b}}){Mowla}, {van Dokkum}, {Brammer},
  {Momcheva}, {van der Wel}, {Whitaker}, {Nelson}, {Bezanson}, {Muzzin},
  {Franx}, {MacKenty}, {Leja}, {Kriek}, \& {Marchesini}}]{2019ApJ...880...57M}
{Mowla}, L.~A., {van Dokkum}, P., {Brammer}, G.~B., {et~al.}
  2019{\natexlab{b}}, \apj, 880, 57, \dodoi{10.3847/1538-4357/ab290a}

\bibitem[{{Nanayakkara} {et~al.}(2022){Nanayakkara}, {Glazebrook}, {Jacobs},
  {Schreiber}, {Brammer}, {Esdaile}, {Kacprzak}, {Labbe}, {Lagos},
  {Marchesini}, {Marsan}, {Nateghi}, {Oesch}, {Papovich}, {Remus}, \&
  {Tran}}]{2022arXiv221211638N}
{Nanayakkara}, T., {Glazebrook}, K., {Jacobs}, C., {et~al.} 2022, arXiv
  e-prints, arXiv:2212.11638, \dodoi{10.48550/arXiv.2212.11638}

\bibitem[{{Nandra} {et~al.}(2015){Nandra}, {Laird}, {Aird}, {Salvato},
  {Georgakakis}, {Barro}, {Perez-Gonzalez}, {Barmby}, {Chary}, {Coil},
  {Cooper}, {Davis}, {Dickinson}, {Faber}, {Fazio}, {Guhathakurta}, {Gwyn},
  {Hsu}, {Huang}, {Ivison}, {Koo}, {Newman}, {Rangel}, {Yamada}, \&
  {Willmer}}]{2015ApJS..220...10N}
{Nandra}, K., {Laird}, E.~S., {Aird}, J.~A., {et~al.} 2015, \apjs, 220, 10,
  \dodoi{10.1088/0067-0049/220/1/10}

\bibitem[{{Nedkova} {et~al.}(2021){Nedkova}, {H{\"a}u{\ss}ler}, {Marchesini},
  {Dimauro}, {Brammer}, {Eigenthaler}, {Feinstein}, {Ferguson},
  {Huertas-Company}, {Johnston}, {Kado-Fong}, {Kartaltepe}, {Labb{\'e}},
  {Lange-Vagle}, {Martis}, {McGrath}, {Muzzin}, {Oesch}, {Ordenes-Brice{\~n}o},
  {Puzia}, {Shipley}, {Simmons}, {Skelton}, {Stefanon}, {van der Wel}, \&
  {Whitaker}}]{2021MNRAS.506..928N}
{Nedkova}, K.~V., {H{\"a}u{\ss}ler}, B., {Marchesini}, D., {et~al.} 2021,
  \mnras, 506, 928, \dodoi{10.1093/mnras/stab1744}

\bibitem[{{Newman} {et~al.}(2018){Newman}, {Belli}, {Ellis}, \&
  {Patel}}]{2018ApJ...862..126N}
{Newman}, A.~B., {Belli}, S., {Ellis}, R.~S., \& {Patel}, S.~G. 2018, \apj,
  862, 126, \dodoi{10.3847/1538-4357/aacd4f}

\bibitem[{Newville {et~al.}(2014)Newville, Stensitzki, Allen, \&
  Ingargiola}]{newville_matthew_2014_11813}
Newville, M., Stensitzki, T., Allen, D.~B., \& Ingargiola, A. 2014, {LMFIT:
  Non-Linear Least-Square Minimization and Curve-Fitting for Python}, 0.8.0,
  Zenodo, \dodoi{10.5281/zenodo.11813}

\bibitem[{{Nipoti} {et~al.}(2012){Nipoti}, {Treu}, {Leauthaud}, {Bundy},
  {Newman}, \& {Auger}}]{2012MNRAS.422.1714N}
{Nipoti}, C., {Treu}, T., {Leauthaud}, A., {et~al.} 2012, \mnras, 422, 1714,
  \dodoi{10.1111/j.1365-2966.2012.20749.x}

\bibitem[{{Oke} \& {Gunn}(1983)}]{1983ApJ...266..713O}
{Oke}, J.~B., \& {Gunn}, J.~E. 1983, \apj, 266, 713, \dodoi{10.1086/160817}

\bibitem[{{Ono} {et~al.}(2022){Ono}, {Harikane}, {Ouchi}, {Yajima}, {Abe},
  {Isobe}, {Shibuya}, {Zhang}, {Nakajima}, \& {Umeda}}]{2022arXiv220813582O}
{Ono}, Y., {Harikane}, Y., {Ouchi}, M., {et~al.} 2022, arXiv e-prints,
  arXiv:2208.13582, \dodoi{10.48550/arXiv.2208.13582}

\bibitem[{{Ormerod} {et~al.}(2024){Ormerod}, {Conselice}, {Adams}, {Harvey},
  {Austin}, {Trussler}, {Ferreira}, {Caruana}, {Lucatelli}, {Li}, \&
  {Roper}}]{2024MNRAS.527.6110O}
{Ormerod}, K., {Conselice}, C.~J., {Adams}, N.~J., {et~al.} 2024, \mnras, 527,
  6110, \dodoi{10.1093/mnras/stad3597}

\bibitem[{{Oser} {et~al.}(2012){Oser}, {Naab}, {Ostriker}, \&
  {Johansson}}]{2012ApJ...744...63O}
{Oser}, L., {Naab}, T., {Ostriker}, J.~P., \& {Johansson}, P.~H. 2012, \apj,
  744, 63, \dodoi{10.1088/0004-637X/744/1/63}

\bibitem[{{Pacifici} {et~al.}(2023){Pacifici}, {Iyer}, {Mobasher}, {da Cunha},
  {Acquaviva}, {Burgarella}, {Calistro Rivera}, {Carnall}, {Chang}, {Chartab},
  {Cooke}, {Fairhurst}, {Kartaltepe}, {Leja}, {Ma{\l}ek}, {Salmon}, {Torelli},
  {Vidal-Garc{\'\i}a}, {Boquien}, {Brammer}, {Brown}, {Capak}, {Chevallard},
  {Circosta}, {Croton}, {Davidzon}, {Dickinson}, {Duncan}, {Faber}, {Ferguson},
  {Fontana}, {Guo}, {Haeussler}, {Hemmati}, {Jafariyazani}, {Kassin}, {Larson},
  {Lee}, {Mantha}, {Marchi}, {Nayyeri}, {Newman}, {Pandya}, {Pforr}, {Reddy},
  {Sanders}, {Shah}, {Shahidi}, {Stevans}, {Triani}, {Tyler}, {Vanderhoof}, {de
  la Vega}, {Wang}, \& {Weston}}]{2023ApJ...944..141P}
{Pacifici}, C., {Iyer}, K.~G., {Mobasher}, B., {et~al.} 2023, \apj, 944, 141,
  \dodoi{10.3847/1538-4357/acacff}

\bibitem[{{Paufique} {et~al.}(2010){Paufique}, {Bruton}, {Glindemann}, {Jost},
  {Kolb}, {Jochum}, {Le Louarn}, {Kiekebusch}, {Hubin}, {Madec}, {Conzelmann},
  {Siebenmorgen}, {Donaldson}, {Arsenault}, \& {Tordo}}]{Paufique2010}
{Paufique}, J., {Bruton}, A., {Glindemann}, A., {et~al.} 2010, in Society of
  Photo-Optical Instrumentation Engineers (SPIE) Conference Series, Vol. 7736,
  Adaptive Optics Systems II, ed. B.~L. {Ellerbroek}, M.~{Hart}, N.~{Hubin}, \&
  P.~L. {Wizinowich}, 77361P, \dodoi{10.1117/12.858261}

\bibitem[{{Peng} {et~al.}(2002){Peng}, {Ho}, {Impey}, \&
  {Rix}}]{2002AJ....124..266P}
{Peng}, C.~Y., {Ho}, L.~C., {Impey}, C.~D., \& {Rix}, H.-W. 2002, \aj, 124,
  266, \dodoi{10.1086/340952}

\bibitem[{{Peng} {et~al.}(2010){Peng}, {Ho}, {Impey}, \&
  {Rix}}]{2010AJ....139.2097P}
---. 2010, \aj, 139, 2097, \dodoi{10.1088/0004-6256/139/6/2097}

\bibitem[{{P{\'e}rez-Gonz{\'a}lez} {et~al.}(2023){P{\'e}rez-Gonz{\'a}lez},
  {Barro}, {Annunziatella}, {Costantin}, {Garc{\'\i}a-Argum{\'a}nez},
  {McGrath}, {M{\'e}rida}, {Zavala}, {Haro}, {Bagley}, {Backhaus}, {Behroozi},
  {Bell}, {Bisigello}, {Buat}, {Calabr{\`o}}, {Casey}, {Cleri}, {Coogan},
  {Cooper}, {Cooray}, {Dekel}, {Dickinson}, {Elbaz}, {Ferguson}, {Finkelstein},
  {Fontana}, {Franco}, {Gardner}, {Giavalisco}, {G{\'o}mez-Guijarro},
  {Grazian}, {Grogin}, {Guo}, {Huertas-Company}, {Jogee}, {Kartaltepe},
  {Kewley}, {Kirkpatrick}, {Kocevski}, {Koekemoer}, {Long}, {Lotz}, {Lucas},
  {Papovich}, {Pirzkal}, {Ravindranath}, {Somerville}, {Tacchella}, {Trump},
  {Wang}, {Wilkins}, {Wuyts}, {Yang}, \& {Yung}}]{2023ApJ...946L..16P}
{P{\'e}rez-Gonz{\'a}lez}, P.~G., {Barro}, G., {Annunziatella}, M., {et~al.}
  2023, \apjl, 946, L16, \dodoi{10.3847/2041-8213/acb3a5}

\bibitem[{{Perrin} {et~al.}(2014){Perrin}, {Sivaramakrishnan}, {Lajoie},
  {Elliott}, {Pueyo}, {Ravindranath}, \& {Albert}}]{2014SPIE.9143E..3XP}
{Perrin}, M.~D., {Sivaramakrishnan}, A., {Lajoie}, C.-P., {et~al.} 2014, in
  Society of Photo-Optical Instrumentation Engineers (SPIE) Conference Series,
  Vol. 9143, Space Telescopes and Instrumentation 2014: Optical, Infrared, and
  Millimeter Wave, ed. J.~{Oschmann}, Jacobus~M., M.~{Clampin}, G.~G. {Fazio},
  \& H.~A. {MacEwen}, 91433X, \dodoi{10.1117/12.2056689}

\bibitem[{{Perrin} {et~al.}(2012){Perrin}, {Soummer}, {Elliott}, {Lallo}, \&
  {Sivaramakrishnan}}]{2012SPIE.8442E..3DP}
{Perrin}, M.~D., {Soummer}, R., {Elliott}, E.~M., {Lallo}, M.~D., \&
  {Sivaramakrishnan}, A. 2012, in Society of Photo-Optical Instrumentation
  Engineers (SPIE) Conference Series, Vol. 8442, Space Telescopes and
  Instrumentation 2012: Optical, Infrared, and Millimeter Wave, ed. M.~C.
  {Clampin}, G.~G. {Fazio}, H.~A. {MacEwen}, \& J.~{Oschmann}, Jacobus~M.,
  84423D, \dodoi{10.1117/12.925230}

\bibitem[{{Pirard} {et~al.}(2004){Pirard}, {Kissler-Patig}, {Moorwood},
  {Biereichel}, {Delabre}, {Dorn}, {Finger}, {Gojak}, {Huster}, {Jung}, {Koch},
  {Le Louarn}, {Lizon}, {Mehrgan}, {Pozna}, {Silber}, {Sokar}, \&
  {Stegmeier}}]{Pirard2004}
{Pirard}, J.-F., {Kissler-Patig}, M., {Moorwood}, A., {et~al.} 2004, in Society
  of Photo-Optical Instrumentation Engineers (SPIE) Conference Series, Vol.
  5492, Ground-based Instrumentation for Astronomy, ed. A.~F.~M. {Moorwood} \&
  M.~{Iye}, 1763--1772, \dodoi{10.1117/12.578293}

\bibitem[{{Pontoppidan} {et~al.}(2022){Pontoppidan}, {Barrientes}, {Blome},
  {Braun}, {Brown}, {Carruthers}, {Coe}, {DePasquale}, {Espinoza}, {Marin},
  {Gordon}, {Henry}, {Hustak}, {James}, {Jenkins}, {Koekemoer}, {LaMassa},
  {Law}, {Lockwood}, {Moro-Martin}, {Mullally}, {Pagan}, {Player}, {Proffitt},
  {Pulliam}, {Ramsay}, {Ravindranath}, {Reid}, {Robberto}, {Sabbi}, {Ubeda},
  {Balogh}, {Flanagan}, {Gardner}, {Hasan}, {Meinke}, \&
  {Nota}}]{2022ApJ...936L..14P}
{Pontoppidan}, K.~M., {Barrientes}, J., {Blome}, C., {et~al.} 2022, \apjl, 936,
  L14, \dodoi{10.3847/2041-8213/ac8a4e}

\bibitem[{Price-Whelan {et~al.}(2018)Price-Whelan, Sip^^c5^^91cz,
  G{\"{u}}nther, Lim, Crawford, Conseil, Shupe, Craig, Dencheva, Ginsburg,
  VanderPlas, Bradley, P{\'{e}}rez-Su{\'{a}}rez, de~Val-Borro, Aldcroft, Cruz,
  Robitaille, Tollerud, Ardelean, Babej, Bach, Bachetti, Bakanov, Bamford,
  Barentsen, Barmby, Baumbach, Berry, Biscani, Boquien, Bostroem, Bouma,
  Brammer, Bray, Breytenbach, Buddelmeijer, Burke, Calderone, Rodr{\'{i}}guez,
  Cara, Cardoso, Cheedella, Copin, Corrales, Crichton, D'Avella, Deil, Depagne,
  Dietrich, Donath, Droettboom, Earl, Erben, Fabbro, Ferreira, Finethy, Fox,
  Garrison, Gibbons, Goldstein, Gommers, Greco, Greenfield, Groener, Grollier,
  Hagen, Hirst, Homeier, Horton, Hosseinzadeh, Hu, Hunkeler, Ivezi{\'{c}},
  Jain, Jenness, Kanarek, Kendrew, Kern, Kerzendorf, Khvalko, King, Kirkby,
  Kulkarni, Kumar, Lee, Lenz, Littlefair, Ma, Macleod, Mastropietro, McCully,
  Montagnac, Morris, Mueller, Mumford, Muna, Murphy, Nelson, Nguyen, Ninan,
  N{\"{o}}the, Ogaz, Oh, Parejko, Parley, Pascual, Patil, Patil, Plunkett,
  Prochaska, Rastogi, Janga, Sabater, Sakurikar, Seifert, Sherbert,
  Sherwood-Taylor, Shih, Sick, Silbiger, Singanamalla, Singer, Sladen, Sooley,
  Sornarajah, Streicher, Teuben, Thomas, Tremblay, Turner, Terr{\'{o}}n, van
  Kerkwijk, de~la Vega, Watkins, Weaver, Whitmore, Woillez, \&
  Zabalza}]{Price-Whelan2018}
Price-Whelan, A.~M., Sip^^c5^^91cz, B.~M., G{\"{u}}nther, H.~M., {et~al.} 2018,
  \aj, 156, 123, \dodoi{10.3847/1538-3881/aabc4f}

\bibitem[{Robitaille {et~al.}(2013)Robitaille, Tollerud, Greenfield,
  Droettboom, Bray, Aldcroft, Davis, Ginsburg, Price-Whelan, Kerzendorf,
  Conley, Crighton, Barbary, Muna, Ferguson, Grollier, Parikh, Nair,
  G{\"{u}}nther, Deil, Woillez, Conseil, Kramer, Turner, Singer, Fox, Weaver,
  Zabalza, Edwards, {Azalee Bostroem}, Burke, Casey, Crawford, Dencheva, Ely,
  Jenness, Labrie, Lim, Pierfederici, Pontzen, Ptak, Refsdal, Servillat, \&
  Streicher}]{Robitaille2013}
Robitaille, T.~P., Tollerud, E.~J., Greenfield, P., {et~al.} 2013, \aap, 558,
  A33, \dodoi{10.1051/0004-6361/201322068}

\bibitem[{{Saracco} {et~al.}(2020){Saracco}, {Marchesini}, {La Barbera},
  {Gargiulo}, {Annunziatella}, {Forrest}, {Lange Vagle}, {Marsan}, {Muzzin},
  {Stefanon}, \& {Wilson}}]{2020ApJ...905...40S}
{Saracco}, P., {Marchesini}, D., {La Barbera}, F., {et~al.} 2020, \apj, 905,
  40, \dodoi{10.3847/1538-4357/abc7c4}

\bibitem[{{Schreiber} {et~al.}(2018){Schreiber}, {Glazebrook}, {Nanayakkara},
  {Kacprzak}, {Labb{\'e}}, {Oesch}, {Yuan}, {Tran}, {Papovich}, {Spitler}, \&
  {Straatman}}]{2018A&A...618A..85S}
{Schreiber}, C., {Glazebrook}, K., {Nanayakkara}, T., {et~al.} 2018, \aap, 618,
  A85, \dodoi{10.1051/0004-6361/201833070}

\bibitem[{{S{\'e}rsic}(1963)}]{1963BAAA....6...41S}
{S{\'e}rsic}, J.~L. 1963, Boletin de la Asociacion Argentina de Astronomia La
  Plata Argentina, 6, 41

\bibitem[{{Shen} {et~al.}(2003){Shen}, {Mo}, {White}, {Blanton}, {Kauffmann},
  {Voges}, {Brinkmann}, \& {Csabai}}]{2003MNRAS.343..978S}
{Shen}, S., {Mo}, H.~J., {White}, S. D.~M., {et~al.} 2003, \mnras, 343, 978,
  \dodoi{10.1046/j.1365-8711.2003.06740.x}

\bibitem[{{Steinhardt} {et~al.}(2021){Steinhardt}, {Jespersen}, \&
  {Linzer}}]{2021ApJ...923....8S}
{Steinhardt}, C.~L., {Jespersen}, C.~K., \& {Linzer}, N.~B. 2021, \apj, 923, 8,
  \dodoi{10.3847/1538-4357/ac2a2f}

\bibitem[{{Straatman} {et~al.}(2015){Straatman}, {Labb{\'e}}, {Spitler},
  {Glazebrook}, {Tomczak}, {Allen}, {Brammer}, {Cowley}, {van Dokkum},
  {Kacprzak}, {Kawinwanichakij}, {Mehrtens}, {Nanayakkara}, {Papovich},
  {Persson}, {Quadri}, {Rees}, {Tilvi}, {Tran}, \&
  {Whitaker}}]{2015ApJ...808L..29S}
{Straatman}, C. M.~S., {Labb{\'e}}, I., {Spitler}, L.~R., {et~al.} 2015, \apjl,
  808, L29, \dodoi{10.1088/2041-8205/808/1/L29}

\bibitem[{{Suess} {et~al.}(2019){Suess}, {Kriek}, {Price}, \&
  {Barro}}]{2019ApJ...877..103S}
{Suess}, K.~A., {Kriek}, M., {Price}, S.~H., \& {Barro}, G. 2019, \apj, 877,
  103, \dodoi{10.3847/1538-4357/ab1bda}

\bibitem[{{Suess} {et~al.}(2021){Suess}, {Kriek}, {Price}, \&
  {Barro}}]{2021ApJ...915...87S}
---. 2021, \apj, 915, 87, \dodoi{10.3847/1538-4357/abf1e4}

\bibitem[{{Suess} {et~al.}(2022){Suess}, {Bezanson}, {Nelson}, {Setton},
  {Price}, {van Dokkum}, {Brammer}, {Labb{\'e}}, {Leja}, {Miller}, {Robertson},
  {Wel}, {Weaver}, \& {Whitaker}}]{2022ApJ...937L..33S}
{Suess}, K.~A., {Bezanson}, R., {Nelson}, E.~J., {et~al.} 2022, \apjl, 937,
  L33, \dodoi{10.3847/2041-8213/ac8e06}

\bibitem[{{Tanaka} {et~al.}(2019){Tanaka}, {Valentino}, {Toft}, {Onodera},
  {Shimakawa}, {Ceverino}, {Faisst}, {Gallazzi}, {G{\'o}mez-Guijarro}, {Kubo},
  {Magdis}, {Steinhardt}, {Stockmann}, {Yabe}, \& {Zabl}}]{2019ApJ...885L..34T}
{Tanaka}, M., {Valentino}, F., {Toft}, S., {et~al.} 2019, \apjl, 885, L34,
  \dodoi{10.3847/2041-8213/ab4ff3}

\bibitem[{{Toft} {et~al.}(2017){Toft}, {Zabl}, {Richard}, {Gallazzi},
  {Zibetti}, {Prescott}, {Grillo}, {Man}, {Lee}, {G{\'o}mez-Guijarro},
  {Stockmann}, {Magdis}, \& {Steinhardt}}]{2017Natur.546..510T}
{Toft}, S., {Zabl}, J., {Richard}, J., {et~al.} 2017, \nat, 546, 510,
  \dodoi{10.1038/nature22388}

\bibitem[{{Tzanavaris} {et~al.}(2014){Tzanavaris}, {Gallagher},
  {Hornschemeier}, {Fedotov}, {Eracleous}, {Brandt}, {Desjardins}, {Charlton},
  \& {Gronwall}}]{2014ApJS..212....9T}
{Tzanavaris}, P., {Gallagher}, S.~C., {Hornschemeier}, A.~E., {et~al.} 2014,
  \apjs, 212, 9, \dodoi{10.1088/0067-0049/212/1/9}

\bibitem[{Valentino(2023)}]{francesco_valentino_2023_7614908}
Valentino, F. 2023, {"An atlas of color-selected quiescent galaxies at z>3 in
  public JWST fields" (Supplementary material)}, 1.0,  Zenodo,
  \dodoi{10.5281/zenodo.7614908}

\bibitem[{{Valentino} {et~al.}(2020){Valentino}, {Tanaka}, {Davidzon}, {Toft},
  {G{\'o}mez-Guijarro}, {Stockmann}, {Onodera}, {Brammer}, {Ceverino},
  {Faisst}, {Gallazzi}, {Hayward}, {Ilbert}, {Kubo}, {Magdis}, {Selsing},
  {Shimakawa}, {Sparre}, {Steinhardt}, {Yabe}, \& {Zabl}}]{2020ApJ...889...93V}
{Valentino}, F., {Tanaka}, M., {Davidzon}, I., {et~al.} 2020, \apj, 889, 93,
  \dodoi{10.3847/1538-4357/ab64dc}

\bibitem[{{Valentino} {et~al.}(2023){Valentino}, {Brammer}, {Gould}, {Kokorev},
  {Fujimoto}, {Jespersen}, {Vijayan}, {Weaver}, {Ito}, {Tanaka}, {Ilbert},
  {Magdis}, {Whitaker}, {Faisst}, {Gallazzi}, {Gillman}, {Gim{\'e}nez-Arteaga},
  {G{\'o}mez-Guijarro}, {Kubo}, {Heintz}, {Hirschmann}, {Oesch}, {Onodera},
  {Rizzo}, {Lee}, {Strait}, \& {Toft}}]{2023ApJ...947...20V}
{Valentino}, F., {Brammer}, G., {Gould}, K. M.~L., {et~al.} 2023, \apj, 947,
  20, \dodoi{10.3847/1538-4357/acbefa}

\bibitem[{{van der Wel} {et~al.}(2012){van der Wel}, {Bell}, {H{\"a}ussler},
  {McGrath}, {Chang}, {Guo}, {McIntosh}, {Rix}, {Barden}, {Cheung}, {Faber},
  {Ferguson}, {Galametz}, {Grogin}, {Hartley}, {Kartaltepe}, {Kocevski},
  {Koekemoer}, {Lotz}, {Mozena}, {Peth}, \& {Peng}}]{2012ApJS..203...24V}
{van der Wel}, A., {Bell}, E.~F., {H{\"a}ussler}, B., {et~al.} 2012, \apjs,
  203, 24, \dodoi{10.1088/0067-0049/203/2/24}

\bibitem[{{van der Wel} {et~al.}(2014){van der Wel}, {Franx}, {van Dokkum},
  {Skelton}, {Momcheva}, {Whitaker}, {Brammer}, {Bell}, {Rix}, {Wuyts},
  {Ferguson}, {Holden}, {Barro}, {Koekemoer}, {Chang}, {McGrath},
  {H{\"a}ussler}, {Dekel}, {Behroozi}, {Fumagalli}, {Leja}, {Lundgren},
  {Maseda}, {Nelson}, {Wake}, {Patel}, {Labb{\'e}}, {Faber}, {Grogin}, \&
  {Kocevski}}]{2014ApJ...788...28V}
{van der Wel}, A., {Franx}, M., {van Dokkum}, P.~G., {et~al.} 2014, \apj, 788,
  28, \dodoi{10.1088/0004-637X/788/1/28}

\bibitem[{{van Dokkum} {et~al.}(2015){van Dokkum}, {Nelson}, {Franx}, {Oesch},
  {Momcheva}, {Brammer}, {F{\"o}rster Schreiber}, {Skelton}, {Whitaker}, {van
  der Wel}, {Bezanson}, {Fumagalli}, {Illingworth}, {Kriek}, {Leja}, \&
  {Wuyts}}]{2015ApJ...813...23V}
{van Dokkum}, P.~G., {Nelson}, E.~J., {Franx}, M., {et~al.} 2015, \apj, 813,
  23, \dodoi{10.1088/0004-637X/813/1/23}

\bibitem[{Virtanen {et~al.}(2020)Virtanen, Gommers, Oliphant, Haberland, Reddy,
  Cournapeau, Burovski, Peterson, Weckesser, Bright, {van der Walt}, Brett,
  Wilson, Millman, Mayorov, Nelson, Jones, Kern, Larson, Carey, Polat, Feng,
  Moore, {VanderPlas}, Laxalde, Perktold, Cimrman, Henriksen, Quintero, Harris,
  Archibald, Ribeiro, Pedregosa, {van Mulbregt}, \& {SciPy 1.0
  Contributors}}]{2020SciPy-NMeth}
Virtanen, P., Gommers, R., Oliphant, T.~E., {et~al.} 2020, Nature Methods, 17,
  261, \dodoi{10.1038/s41592-019-0686-2}

\bibitem[{{Ward} {et~al.}(2023){Ward}, {de la Vega}, {Mobasher}, {McGrath},
  {Iyer}, {Calabro}, {Costantin}, {Dickinson}, {Holwerda}, {Huertas-Company},
  {Hirschmann}, {Lucas}, {Pandya}, {Wilkins}, {Yung}, {Arrabal Haro}, {Bagley},
  {Finkelstein}, {Kartaltepe}, {Koekemoer}, {Papovich}, \&
  {Pirzkal}}]{2023arXiv231102162W}
{Ward}, E.~M., {de la Vega}, A., {Mobasher}, B., {et~al.} 2023, arXiv e-prints,
  arXiv:2311.02162, \dodoi{10.48550/arXiv.2311.02162}

\bibitem[{{Weaver} {et~al.}(2022){Weaver}, {Davidzon}, {Toft}, {Ilbert},
  {McCracken}, {Gould}, {Jespersen}, {Steinhardt}, {Lagos}, {Capak}, {Casey},
  {Chartab}, {Faisst}, {Hayward}, {Kartaltepe}, {Kauffmann}, {Koekemoer},
  {Kokorev}, {Laigle}, {Liu}, {Long}, {Magdis}, {McPartland}, {Milvang-Jensen},
  {Mobasher}, {Moneti}, {Peng}, {Sanders}, {Shuntov}, {Sneppen}, {Valentino},
  {Zalesky}, \& {Zamorani}}]{2022arXiv221202512W}
{Weaver}, J.~R., {Davidzon}, I., {Toft}, S., {et~al.} 2022, arXiv e-prints,
  arXiv:2212.02512, \dodoi{10.48550/arXiv.2212.02512}

\bibitem[{{Williams} {et~al.}(2009){Williams}, {Quadri}, {Franx}, {van Dokkum},
  \& {Labb{\'e}}}]{2009ApJ...691.1879W}
{Williams}, R.~J., {Quadri}, R.~F., {Franx}, M., {van Dokkum}, P., \&
  {Labb{\'e}}, I. 2009, \apj, 691, 1879, \dodoi{10.1088/0004-637X/691/2/1879}

\bibitem[{{Yang} {et~al.}(2022){Yang}, {Morishita}, {Leethochawalit},
  {Castellano}, {Calabr{\`o}}, {Treu}, {Bonchi}, {Fontana}, {Mason}, {Merlin},
  {Paris}, {Trenti}, {Roberts-Borsani}, {Bradac}, {Vanzella}, {Vulcani},
  {Marchesini}, {Ding}, {Nanayakkara}, {Birrer}, {Glazebrook}, {Jones},
  {Boyett}, {Santini}, {Strait}, \& {Wang}}]{2022ApJ...938L..17Y}
{Yang}, L., {Morishita}, T., {Leethochawalit}, N., {et~al.} 2022, \apjl, 938,
  L17, \dodoi{10.3847/2041-8213/ac8803}

\bibitem[{{Zhuang} \& {Shen}(2023)}]{2023arXiv230413776Z}
{Zhuang}, M.-Y., \& {Shen}, Y. 2023, arXiv e-prints, arXiv:2304.13776,
  \dodoi{10.48550/arXiv.2304.13776}

\end{thebibliography}

\end{document}